\documentclass[aps,preprint]{revtex4}
\usepackage{amssymb}
\usepackage{amsmath}
\usepackage{graphicx}

\begin{document}

\title{Scaling and Formulary cross sections for ion-atom impact
ionization}
\author{Igor D. Kaganovich, Edward Startsev and Ronald C. Davidson}
\affiliation{Plasma Physics Laboratory, Princeton University, Princeton, NJ 08543}
\date{\today}

\begin{abstract}
The values of ion-atom ionization cross sections are frequently needed for
many applications that utilize the propagation of fast ions through matter.
When experimental data and theoretical calculations are not available,
approximate formulas are frequently used. This paper briefly summarizes the
most important theoretical results and approaches to cross section
calculations in order to place the discussion in historical perspective and
offer a concise introduction to the topic. Based on experimental data and
theoretical predictions, a new fit for ionization cross sections is
proposed. The range of validity and accuracy of several frequently used
approximations (classical trajectory, the Born approximation, and so forth)
are discussed using, as examples, the ionization cross sections of hydrogen
and helium atoms by various fully stripped ions.
\end{abstract}

\maketitle

\section{Introduction}

Ion-atom ionizing collisions play an important role in many applications
such as heavy ion inertial fusion \cite{HIF reference}, collisional and
radiative processes in the Earth's upper atmosphere \cite{atmosphere},
ion-beam lifetimes in accelerators \cite{accelerators life time}, atomic
spectroscopy \cite{spectroscopy}, and ion stopping in matter \cite{beam
stopping}, and are of considerable interest in atomic physics \cite{Review
atomic physics}. The recent resurgence of interest in charged particle beam
transport in background plasma is brought about by the\ recognition that
plasma can be used as a magnetic lens. Applications of the plasma lens
ranging from heavy ion fusion to high energy lepton colliders are discussed
in Refs. [6-10]. In particular, both heavy ion fusion and high energy
physics applications involve the transport of \textit{positive} charges in
plasma: partially stripped heavy elements for heavy ion fusion; positrons
for electron-positrons colliders \cite{hep lens}; and high-density
laser-produced proton beams for the fast ignition of inertial confinement
fusion targets \cite{ICF Fast Ignitor}.

To estimate the ionization and stripping rates of fast ions
propagating through gas or plasma, the values of ion-atom
ionization cross sections are necessary. In contrast to the
electron \cite{Voronov} and proton \cite{Rudd, Rudd 2, Ogurtsov}
ionization cross sections, where experimental data or theoretical
calculations exist for practically any ion and atom, the knowledge
of ionization cross sections by fast complex ions and atoms is far
from complete \cite{Shvelko book, Dowell1, Dowell2,Daniel}. When
experimental data and theoretical calculations are not available,
approximate formulas are frequently used.

The raison d'etre for this paper are the frequent requests that we have had
from colleagues for a paper describing the regions of validity of different
approximations and scaling laws in the calculation of ion- atom stripping
cross sections. The experimental data on stripping cross sections at low
projectile energy were collected in the late 1980s, while comprehensive
quantum mechanical simulations were performed in the late 1990s. Having in
hand both new experimental data and simulation results enabled us to
identify regions of validity of different approximations and propose a new
scaling law, which is the subject of the present paper.

The most popular formula for ionization cross sections was proposed by
Gryzinski \cite{Gryz}. The web of science search engine \cite{webofscience}
shows 457 citations of the paper, and most of the citing papers use
Gryzinski's formula to evaluate the cross sections. In this approach, the
cross section is specified by multiplication of a scaling factor and the
unique function of the projectile velocity normalized to the orbital
electron velocity. The popularity of Gryzinski's formula is based on the
simplicity of the calculation, notwithstanding the fact that his formula is
not accurate at small energies.

Another fit, proposed by Gillespie, gives results close to Gryzinski's
formula at large energies, and makes corrections to Gryzinski's formula at
small energies \cite{Gellipsie}. Although more accurate, Gillespie's fit is
not frequently used in applications, because it requires a knowledge of
fitting parameters not always known a priori.

In this paper, we propose a new fit formula for ionization cross section
which has no fitting parameters. The formula is checked against available
experimental data and theoretical predictions. Note that previous scaling
laws either used fitting parameters or actually did not match experiments
for a wide range of projectile velocities. We also briefly review the most
important theoretical results and approaches to cross section calculations
in order to place the discussion in historical perspective and offer
nonspecialists a concise introduction to the topic.

The organization of this paper is as follows. In Sec.II we give a brief
overview of key theoretical results and experimental data. Further details
of the theoretical models are presented in Appendices A-C. The new proposed
fit formula for ionization cross section is presented in Sec.III, including
a detailed comparison with experimental data, and in Sec.IV the theoretical
justification for the new fit formula is discussed.

\section{Brief Overview of the theoretical models and experimental data for
ionization cross sections}

There are several theoretical approaches to cross section
calculations. These include: classical calculations that make use
of a classical trajectory and the atomic electron velocity
distribution functions given by quantum mechanics [this approach
is frequently referred to as classical trajectory Monte Carlo
(CTMC)]; quantum mechanical calculations based on Born, eikonal or
quasiclassical approximations, and so forth \cite{Shvelko book,
Dowell1, Dowell2, Daniel}. All approaches are computationally
intensive and the error and range of validity are difficult to
estimate in most cases. Therefore, different fittings and scalings
for cross sections are frequently used in practical applications.

Most scalings were developed using theories and simulations based
on classical mechanics. Classical trajectory calculations are
easier to perform compared with quantum mechanical calculations.
Moreover, in some cases the CTMC calculations yield results very
close to the quantum-mechanical calculations \cite{our PoP hif,
Olson exp, Watson exp, Mueller new}. The reason for similar
results lies in the fact that the Rutherford scattering cross
section is identical in both classical and quantum-mechanical
derivations \cite{Landau book}. Therefore, when an ionizing
collision is predominantly a consequence of the electron
scattering at small impact parameters close to the nucleus, the
quantum mechanical uncertainty in the scattering angle is small
compared with the angle itself, and the classical calculation can
yield an accurate description \cite{Bohr, my PAC Xsection, my
HIF}. Whereas in the opposite limit, when an ionizing collision is
predominantly a consequence of the electron scattering at large
impact parameters far from the nucleus, the quantum mechanical
uncertainty in the scattering angle is large compared with the
angle itself, and the classical calculation can remarkably fail in
computing the ionization cross section \cite{myPRA, Japan exp}.

In the present analysis, we consider first the stripping or ionization cross
section of the hydrogen-like electron orbits (for example one-electron
ions), with nucleus of charge $Z_{T},$ colliding with a fully stripped ion
of charge $Z_{p}$. Subsequently, we show that the approach can be
generalized with reasonable accuracy for any electron orbital, making use of
the ionization potential of the electron orbitals. Because different
terminology is used in the literature, we call a \textit{stripping collision}
a collision in which the fast ion loses an electron in a collision with a
stationary target ion or atom (in the laboratory frame); and we call an
\textit{ionizing collision } a collision in which a fast ion ionizes a
stationary target ion or atom \cite{Shvelko book}. Both cases are physically
equivalent to each other by changing the frame of reference, and further
consideration can be given in the frame of the atom or ion being ionized.

Atomic units are used throughout this paper with $e=\hbar =m_{e}=1$, which
corresponds to length normalized to $a_{0}=\hbar
^{2}/(m_{e}e^{2})=0.529\cdot 10^{-8}cm,$ velocity normalized to $%
v_{0}=e^{2}/\hbar =2.19\cdot 10^{8}cm/s$, energy normalized to $%
E_{0}=m_{e}v_{0}^{2}=2Ry=27.2eV$, where $Ry$ is the Rydberg energy. The
normalizing coefficients are kept in all equations for robust application of
the formulas. For efficient manipulation of the formulas it is worth noting
that the normalized projectile ion velocity is $v/v_{0}=0.2\sqrt{E[keV/amu]}$%
, where $E$ is energy per nucleon in $keV/amu$. Therefore, $25keV/amu$
corresponds to the atomic velocity scale. Some papers express the normalized
velocity $v/v_{0}$ as $\beta \alpha $, where $\beta =v/c$, and $%
v_{0}/c=\alpha =1/137$. Here, $c $ is the speed of light, and $\alpha $ is
the fine structure constant.

For a one-electron ion, the typical scale for the electron orbital velocity
is $v_{nl}=v_{0}Z_{T}$. Here, $n,l$ is the standard notation for the main
quantum number and the orbital angular momentum \cite{Landau book}. The
collision dynamics is very different depending on whether $v$ is smaller or
larger than $v_{nl}$.

\subsection{Behavior of cross sections at large projectile velocities $%
v>v_{nl}$}

If $v>>v_{nl}$, the electron interaction with the projectile ion occurs for
a very short time and the interaction time decreases as the velocity
increases. Therefore, the ionization cross section also decreases as the
velocity increases. In the opposite case $v<<v_{nl}$, the electron
circulation around the target nucleus is much faster than the interaction
time, and the momentum transfer from the projectile ion to the electron
averages out due to the fast circulation. Thus, the cross section decreases
as the projectile velocity decreases. This is why the cross section
typically has a maximum at $v=v_{\max }\sim v_{nl}$, but as we shall see
below, $v_{\max }$ also depends on the charge of the projectile.

\subsubsection{Thompson's treatment}

In the first treatment, Thompson calculated the ionization cross section in
the limit $v>>v_{nl}$ \cite{Thompson}. This treatment neglected completely
the orbital motion of the target electrons and assumed a straight-line
trajectory of the projectile. In this approximation, the velocity kick
acquired by the electron during the collision is entirely in the direction
perpendicular to the ion trajectory, because the final action of the force
along the trajectory cancels out due to symmetry, i.e., the electron
velocity change during the approaching phase is equal to minus the electron
velocity change during the departing phase. The momentum acquired by the
electron ( $m_{e}\Delta v$) from passing by the projectile moving with the
speed $v$ and impact parameter $\rho $ is given by the integral over time of
the force perpendicular to ion trajectory $F_{\perp }=e^{2}Z_{p}\rho /(\rho
^{2}+v^{2}t^{2})^{3/2}$, where $t=0$ corresponds to the distance of the
closest approach. Time integration of the force yields
\begin{equation}
\Delta v(\rho )=\frac{2e^{2}Z_{p}}{m_{e}v\rho }.  \label{velocity kick}
\end{equation}%
From Eq.(\ref{velocity kick}) it follows that only collisions with
sufficiently small impact parameters result in ionization. The minimum
impact parameter for ionization of an initially stationary electron ($\rho
_{\min }$) is $m_{e}\Delta v(\rho _{\min })^{2}/2=I_{nl}$. During a
collision with impact parameter $\rho _{\min }$ the energy transfer from the
projectile to the electron is equal to the ionization potential $%
I_{nl}=Z_{T}^{2}E_{0}/2$, or $\Delta v(\rho _{\min })=v_{nl}$. Substitution
of Eq.(\ref{velocity kick}) gives the total ionization cross section $\pi
\rho _{\min }^{2}$ \cite{Bohr, Thompson}%
\begin{equation}
\sigma ^{Bohr}(v,I_{nl},Z_{p})=2\pi Z_{p}^{2}a_{0}^{2}\,\,\frac{%
v_{0}^{2}E_{0}}{v^{2}I_{nl}}.  \label{Bohr formula}
\end{equation}%
Similarly, Eq.(\ref{Bohr formula}) can be derived by averaging the
Rutherford cross section over all scattering angles leading to ionization.
Although the first derivation of Eq.(\ref{Bohr formula}) was done by
Thompson \cite{Thompson} the formula is frequently referred to as the Bohr
formula \cite{Shvelko book}.

\subsubsection{Gerjuoy's treatment}

The following treatments account for the effect of finite electron orbital
velocity. The most complete and accurate calculations were done by Gerjuoy
\cite{Gerjuoy}. He calculated the differential cross section $d\sigma
/d\Delta E(v_{e},v,\Delta E)$ of energy transfer $\Delta E$\ in the
collision between the projectile ion and a free electron (the target atomic
potential was neglected) with given initial speed $v_{e}$ (and arbitrary
direction), by averaging the Rutherford cross section over all orientations
of electron orbital velocity $\mathbf{v}_{e}$. The total cross section is
then calculated by integration over the energy transfer for energies larger
than the ionization potential, and weighted by the electron velocity
distribution function $f\left( v_{e}\right) $. This gives%
\begin{equation}
\sigma (v,I_{nl},Z_{p})=Z_{p}^{2}\int_{0}^{\infty }\sigma
_{I_{nl}}(v,v_{e})\,f\left( v_{e}\right) \,\,dv_{e},  \label{Gerjuoy formula}
\end{equation}%
where
\begin{equation}
\sigma _{I_{nl}}(v,v_{e})=\int_{I_{nl}}^{\infty }\frac{d\sigma }{d\Delta E}%
(v,v_{e},\Delta E)d\Delta E.  \label{integral dsigma}
\end{equation}%
A rather complicated analytical expression for $d\sigma /d\Delta
E(v_{e},v,\Delta E)$\ is given in Appendix A. For large projectile ion
velocities ($v>>v_{nl}$), the differential cross section can be expressed as
\cite{Gerjuoy}%
\begin{equation}
\frac{d\sigma _{classical}^{high-energy}}{d\Delta E}(v,v_{e},\Delta E)=2\pi
a_{0}^{2}\,\,\frac{E_{0}^{2}}{\Delta E^{3}m_{e}v^{2}}\left( \frac{%
2m_{e}v_{e}^{2}}{3}+\Delta E\right) .
\label{Gerjuoy differential large velocity}
\end{equation}%
Substituting Eq.(\ref{Gerjuoy differential large velocity}) into Eq.(\ref%
{Gerjuoy formula}) and Eq.(\ref{integral dsigma}) gives
\begin{equation}
\sigma _{classical}^{high-energy}(v,I_{nl},Z_{p})=\frac{5}{3}B_{nl}\sigma
^{Bohr}(v,I_{nl},Z_{p}),  \label{high energy Limit}
\end{equation}%
\begin{equation}
B_{nl}\equiv \frac{3}{5}\left( \frac{2K_{nl}}{3I_{nl}}+1\right) ,
\label{Bnl}
\end{equation}%
where $\sigma ^{Bohr}$is given by Eq.(\ref{Bohr formula}), and $K_{nl}\equiv
<m_{e}v_{e}^{2}/2>_{nl}$ is the average orbital electron kinetic energy. For
hydrogen-like electron orbitals, the average electron kinetic energy is
equal to the ionization potential $K_{nl}=I_{nl}$ \cite{Landau book}, and $%
B_{nl}=1$. The $B_{nl}$ factors are introduced to account for the difference
in the electron velocity distribution functions (EVDF) from the EVDF of the
hydrogen-like electron orbitals. The data for $K_{nl}$ are calculated for
many atoms in Ref. \cite{Ponce}. For example, the average kinetic energy for
the helium atom is $K_{nl}\equiv <m_{e}v_{e}^{2}/2>=1.43E_{0}$, whereas $%
I_{nl}=0.91E_{0}$, and therefore $B_{He}=1.22$. That is the reason that
accounting for the finite orbital electron velocity gives a cross section
which is $5/3$ times larger than the Bohr formula in Eq.(\ref{Bohr formula}%
). This is a consequence of the fact that for an electron with nonzero
velocity less energy transfer is required for ionization.

Classical mechanics gives the EVDF as a microcanonical ensemble, where
\begin{equation*}
f\left( v_{e}\right) =Cv_{e}^{2}\int \delta \left( \frac{m_{e}v_{e}^{2}}{2}-%
\frac{Z_{T}}{r}+I_{nl}\right) r^{2}dr.
\end{equation*}%
Here, $C$ is a normalization constant defined so that $\int \,f\left(
v_{e}\right) dv_{e}=1$, and $\delta (...)$ denotes the Dirac delta-function.
Interestingly, the EVDF for a one-electron ion is identical in both the
quantum-mechanical and classical calculations \cite{Landau book, Ponce} with
\begin{equation}
\,f\left( v_{e}\right) \,=\frac{32v_{nl}^{5}}{\pi }\frac{v_{e}^{2}}{\left[
v_{e}^{2}+v_{nl}^{2}\right] ^{4}},  \label{EVDF}
\end{equation}%
where $v_{nl}$ is the scale of electron orbital velocity
\begin{equation}
v_{nl}=v_{0}\sqrt{2I_{nl}/E_{0}}.  \label{vnl on Inl}
\end{equation}%
Although a microcanonical distribution provides the same velocity
distribution as in quantum theory for hydrogen-like shells, this
is not the case for other electron shells. Moreover, the spatial
distribution of the charge density is poorly approximated even for
hydrogen, vanishing
identically for $r>2a_{0}$ rather than decreasing exponentially \cite%
{Dowell2}.  Substituting the general differential cross section
$d\sigma /d\Delta E(v_{e},v,\Delta E)$\ from Eq.(\ref{Gerjuoy
differential}) of Appendix A and the EVDF in Eq.(\ref{EVDF}) into
Eq.(\ref{Gerjuoy formula}) yields
\begin{equation}
\sigma ^{GGV}(v,I_{nl},Z_{p})=\pi a_{0}^{2}Z_{p}^{2}\,\,\frac{E_{0}^{2}}{%
I_{nl}^{2}}G^{GGV}\left( \frac{v}{v_{nl}}\right) .
\label{final classic result}
\end{equation}%
Here, the scaling function $G^{GGV}(x)$ is given by Eq.(\ref{G(x) classical}%
) in Appendix A, using the tabulation of the function $G(x)$\ presented in
Ref.\cite{Vriens} for $x>1$, and in Ref.\cite{Armel thesis} for $x<1$. The
notation GGV stands for the classical trajectory calculation in Eq.(\ref%
{G(x) classical}) due to Gerjuoy \cite{Gerjuoy} using the fit of Garcia and
Vriens \cite{Vriens}.

\subsubsection{Bethe's treatment}

The classical calculations underestimate the cross sections for very high
projectile velocities $v>>v_{nl}$. The scattering angle of the projectile
due to collision with the target atom is of order $\theta _{c}=\Delta p/Mv$,
where $\Delta p$ is the momentum transfer in the collision, and $M$ is the
mass of the projectile particle. The minimum energy transfer from the
projectile is determined by the ionization potential, with $\Delta E=v\Delta
p>I_{nl}$, and $\Delta p>\Delta p_{\min }\equiv I_{nl}/v$. Here, we use the
fact that the momentum transfer $\Delta p$ is predominantly in the direction
perpendicular to the projectile velocity. The projectile particle with wave
vector $k=Mv/\hbar $ undergoes diffraction on the object of the target
atomic size $a_{nl}$ with the diffraction angle of order $\theta
_{d}=1/(ka_{nl})=\hbar /(Mva_{nl})$ \cite{Bohr}. At large projectile
velocities $v>>v_{nl}$, it follows that $\Delta p_{\min }\equiv
I_{nl}/v<<\hbar /a_{nl}$, because $v_{nl}=I_{nl}a_{nl}/\hbar $ for
hydrogen-like electron orbitals. And for small $\Delta p\sim \Delta p_{\min
} $, it follows that $\theta _{c}=\Delta p/Mv<<\theta _{d}=\hbar /(Mva_{nl})$%
. Therefore, the collision can not be described by classical mechanics.

Bethe made use of the Born approximation of quantum mechanics to calculate
cross sections \cite{Bethe} (see Appendix B for details). This yields for $%
v>>v_{nl}$%
\begin{equation}
\sigma ^{Bethe}=\sigma ^{Bohr}(v,I_{nl},Z_{p})\left[ 0.566\ln \left( \frac{v%
}{v_{nl}}\right) +1.261\right] .  \label{Bethe equation}
\end{equation}%
If the projectile speed is much larger than the electron orbital velocity $%
v>>v_{nl}$, the logarithmic term on the right-hand side of Eq.(\ref{Bethe
equation}) contributes substantially to the cross section, and as a result
the quantum mechanical calculation in Eq.(\ref{Bethe equation}) gives a
larger cross section than the classical trajectory treatment in Eq.(\ref%
{high energy Limit}). The quantum mechanical cross section is larger than
the classical trajectory cross section due to the contribution of large
impact parameters ($\rho $) to the quantum-mechanical cross section, where
the ionization is forbidden in classical mechanics because the energy
transfer calculated by classical mechanics is less than the ionization
potential [$\Delta E=v\Delta p_{c}(\rho )<I_{nl}$, where $\Delta p_{c}$ is
the momentum transfer given by classical mechanics in Eq.(\ref{velocity kick}%
)]. However, ionization is possible due to diffraction in quantum mechanics
\cite{Bethe book}. Moreover, integration over these large impact parameters
where the ionization is forbidden in classical mechanics, contributes
considerably to the total ionization cross section (see Appendix B for
further details).

\subsubsection{Gryzinski's treatment}

Gryzinski attempted to obtain the ionization cross sections using only
classical mechanics similarly to Gerjuoy. But, in order to match the
asymptotic behavior of the Bethe formula in Eq.(\ref{Bethe equation}) at
large projectile velocities, Gryzinski assumed an artificial electron
velocity distribution function (EVDF) instead of the correct EVDF in Eq.(\ref%
{EVDF}) \cite{Gryz}, i.e.,%
\begin{equation}
\,f^{Gryz}\left( v_{e}\right) \,=\frac{1}{v_{nl}}\left( \frac{v_{nl}}{v_{e}}%
\right) ^{3}\exp \left( -\frac{v_{nl}}{v_{e}}\right) .  \label{Gryz edf}
\end{equation}%
The ionization cross section was calculated by averaging the Rutherford
cross section over all possible electron velocities, similar to the Gerjuoy
calculation in Eq.(\ref{Gerjuoy formula}), but was less accurate for small
velocities $v<v_{nl}$. The effect of using the EVDF in Eq. (\ref{Gryz edf})
is to populate the EVDF tail with a much larger fraction of high-energy
electrons with $v_{e}>>v_{nl}$, which gives $f^{Gryz}\left( v_{e}\right)
\,\sim v_{e}^{-3}$\ instead of $\,f\left( v_{e}\right) \,\sim v_{e}^{-6}$
for the correct EVDF in Eq.(\ref{EVDF}). As a result, the average electron
kinetic energy $<m_{e}v_{e}^{2}/2>$ diverges, which leads to a considerable
enhancement of the ionization cross section at high projectile velocities.
For $v>>v_{nl}$, Gerjuoy's calculation of the differential cross section $%
d\sigma /d\Delta E(v_{e},v,\Delta E)$ of energy transfer $\Delta E$ is
similar to Gryzinski's. Therefore, we can substitute Eq.(\ref{Gryz edf})
into Eqs.(\ref{Gerjuoy differential large velocity}) and (\ref{integral
dsigma}). Because in the limit $v>>v_{nl}$ the ionization cross section is
proportional to the average electron kinetic energy $<m_{e}v_{e}^{2}/2>$
[Eq.(\ref{high energy Limit})], and the average kinetic energy diverges, it
follows that a small population of high-speed electrons contributes
considerably to the cross section. Using the general expression for $d\sigma
/d\Delta E(v_{e},v,\Delta E)$ avoids singularity and yields the logarithmic
term in the ionization cross section similar to the Bethe formula in Eq.(\ref%
{Bethe equation}). After a number of additional simplifications and
assumptions, Gryzinski suggested an approximation for the cross section in
the form given by Eq.(\ref{final classic result}) with \cite{Gryz}
\begin{equation}
\sigma ^{Gryz}(v,I_{nl},Z_{p})=\pi a_{0}^{2}Z_{p}^{2}\,\,\frac{E_{0}^{2}}{%
I_{nl}^{2}}G^{Gryz}\left( \frac{v}{v_{nl}}\right) .  \label{Gruzinsky}
\end{equation}%
Here, the function $G^{Gryz}(x)$ is specified by Eq.(\ref{Gryzinski G(x)})
of Appendix C. In Eq.(\ref{Gruzinsky}), the function $G^{Gryz}(x)$ has the
following limit
\begin{equation}
G^{Gryz}(x)\rightarrow \left[ 1+0.667\ln (2.7+x)\right] /x^{2}\
\;as\;x\rightarrow \infty ,  \label{Gryzinski G(x) large x}
\end{equation}%
which is close to Bethe's result in Eq.(\ref{Bethe equation}),
\begin{equation}
G^{Bethe}(x)\rightarrow \left[ 1.261+0.566\ln (x)\right] /x^{2}\;as\;x%
\rightarrow \infty .  \label{G bethe}
\end{equation}%
For $10<x<40$, it follows that%
\begin{equation}
G^{Gryz}(x)/G^{Bethe}(x)\simeq 1.04.  \label{Bethe/Gryzinsky}
\end{equation}%
Therefore, the Gryzinski formula can be viewed as a fit to the Bethe formula
at large velocities $v>>v_{nl}$ with some rather arbitrary continuation to
small velocities $v<<v_{nl}$.

\begin{figure}[tbp]
\includegraphics{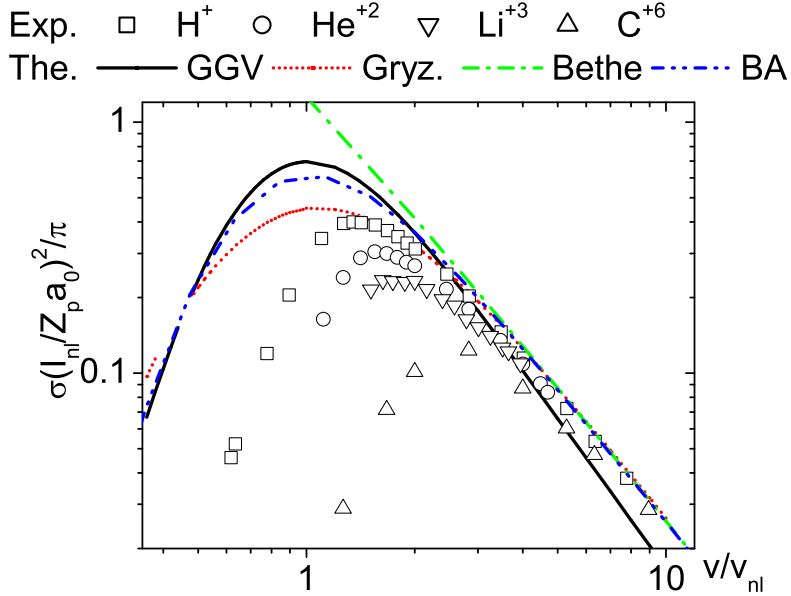}
\caption{Ionization cross sections of hydrogen by fully stripped ions
showing both experimental data and theoretical fits. GGV stands for the
classical calculation in Eq.(\protect\ref{G(x) classical}) due to Gerjuoy
using the fit of Garcia and Vriens. Gryz. denotes the Gryzinski
approximation in Eq.(\protect\ref{Gryzinski G(x)}). Bethe stands for Bethe's
quantum-mechanical calculation in the Born approximation, limited to $%
v>v_{nl}$ in Eq.(\protect\ref{Bethe equation}). Finally, BA denotes the Born
approximation in the general case in Eq.(\protect\ref{BA our fit paper}).
All values are in atomic units. All values are in atomic units. For
hydrogen, the ionization potential is $I_{nl}=1/2E_{0}$, $%
v_{nl}=v_{0}=2.19\cdot 10^{8}cm/s$, and the cross section is normalized to $%
\protect\pi a_{0}^{2}/I_{nl}^{2}=3.51\cdot 10^{-16}cm^{2}$. Symbols show
experimental data.}
\label{Fig.1}
\end{figure}

Figure \ref{Fig.1} shows the experimental data for the cross section for
ionizing collisions of fully stripped ions colliding with a hydrogen atom,
\begin{equation}
X^{q+}+H(1s)\rightarrow X^{q+}+H^{+}+e,  \label{ionization reaction}
\end{equation}%
where $X^{q+}$ denotes fully stripped ions of $H,He,Li,C$ atoms, and ($1s$)
symbolizes the ground state of a hydrogen atom. The experimental data for $%
H^{+}$ ions were taken from \cite{Shah} (note that authors of this reference
concluded that the previous measurements of the cross sections were
inaccurate); from \cite{red books} for $He^{+2},\;C^{+6}$ ions ; and from
\cite{Shah Li atoms} for $Li^{+3}$ ions.

From Fig.\ref{Fig.1} it is evident that the Bethe formula describes well the
cross sections for projectile velocities larger than the orbital velocities $%
v>>v_{nl}$. At large energies, the GGV formula underestimates the cross
sections as discussed before, whereas Gryzinski's formula gives results
close to the Bethe formula and the experimental data. Both, the GGV and
Gryzinski formulas disagree with the experimental data at small energies.

\subsection{Behavior of cross sections at small projectile velocities $%
v<v_{nl}$}

The Bethe, GGV and Gryzinski's formulas fail at small velocities because
they assume free electrons, neglecting the influence of the target atom
potential on the electron motion during the collision. Apparently the
assumption of free electron motion fails if the circulation period of the
electron around the atom's nucleus is comparable with the interaction time
of an ion with the electron. Let us now estimate the projectile velocities
at which the electron circulation needs to be taken into account. The
typical impact parameter leading to ionization is
\begin{equation}
\rho _{ioniz}\simeq \sqrt{\frac{\sigma ^{Bohr}}{\pi }}=\frac{%
2a_{0}v_{0}^{2}Z_{p}}{vv_{nl}},  \label{ionization impact parameter}
\end{equation}%
and the interaction time is of order $\rho _{ioniz}/v$. The electron
circulation time is $\tau _{nl}\simeq a_{nl}/v_{nl}$, where $v_{nl}$ is the
electron orbital velocity, which scales as $v_{nl}=Z_{T}v_{0}$, and $a_{nl}$
is the ion radius $a_{nl}=a_{0}/Z_{T}$ \cite{Bethe book}. Therefore the
condition $\tau _{nl}>\rho _{ioniz}/v$ holds for $v>v_{\max }$, where
\begin{equation}
v_{\max }=v_{nl}\sqrt{2Z_{p}/Z_{T}}.  \label{Maximum cross-section}
\end{equation}%
Here, $Z_{p}$ is the charge of the fully stripped projectile and $Z_{T}$ is
the nuclear charge of the target atom or ion. For velocities larger than $%
v_{\max }$, the ionization cross section decreases as the velocity increases
[see Eq.(\ref{Bethe equation})] due to the decreasing interaction time with
an increase in velocity. On the other hand, for velocities less than $%
v_{\max }$, the collision becomes more adiabatic. The influence of the
projectile is averaged out due to the slower motion of the projectile
compared with the electron orbital velocity, and the ionization cross
section decreases with decreasing projectile velocity. Thus, the cross
section has a maximum at $v\simeq v_{\max }$ [Eq.(\ref{Maximum cross-section}%
)].

Note that if the projectile speed is comparable with or smaller than the
electron orbital velocity $v<v_{nl}$, the Born approximation of quantum
mechanical theory is not valid. Cumbersome quantum mechanical simulations
are necessary for an exact calculation of the cross sections, as for example
in Ref. \cite{Alice paper}. Nevertheless for the case $2Z_{p}\sim Z_{T}$ the
maximum of the cross section calculated from the Born approximation is
similar to the experimental results. To describe the behavior of the cross
section near the maximum, the second-order correction in the parameter $%
v_{nl}/v$ has been calculated in \cite{Kim}, yielding the cross section in
the form%
\begin{equation}
\sigma _{mod}^{Bethe}(\widetilde{v})=\frac{\pi a_{0}^{2}}{I_{nl}^{2}}\,\,%
\frac{Z_{p}^{2}}{\widetilde{v}^{2}}\left[ 0.566\ln \left( \widetilde{v}%
\right) +1.26-0.66\frac{1}{\widetilde{v}^{2}}\right] ,
\label{Bethe with Kim correction}
\end{equation}%
where $\widetilde{v}=v/v_{nl}$. Equation(\ref{Bethe with Kim correction})
agrees with the exact calculation in the Born approximation [Eq.(\ref{BA
cross section})] as described in Appendix B (the agreement is within 10\%
for $\widetilde{v}>1.1$). We have developed the following fit for the cross
section in the Born approximation in the general case,
\begin{equation}
\sigma _{fit}^{BA}(\widetilde{v})=\frac{\pi a_{0}^{2}}{I_{nl}^{2}}\,\,\frac{%
Z_{p}^{2}}{\widetilde{v}^{2}}\left[ 0.283\ln \left( \widetilde{v}%
^{2}+1\right) +1.26\right] \exp \left[ -\frac{1.95}{\widetilde{v}(1+1.2%
\widetilde{v}^{2})}\right] .  \label{BA our fit paper}
\end{equation}%
Equation (\ref{BA our fit paper}) agrees with the exact calculation [Eq.(\ref%
{BA cross section})] within $2\%$ for $\widetilde{v}>1$, and within $20\%$
for $0.2<\widetilde{v}<1.$

Equation (\ref{BA our fit paper}) was derived making use of the unperturbed
atomic electron wave functions, which implicitly assumes that the projectile
particle transfers momentum to the electron and departs to large distances,
where it does not affect the electron to be ionized. The wave function can
therefore be described as a continuous spectrum of the atomic electron, not
affected by the projectile.

This assumption breaks down at low projectile velocities when the projectile
velocity is comparable with the electron orbital velocity. Indeed, the
electron kinetic energy in the frame of the projectile is of order $%
m_{e}v^{2}/2$ and the potential energy $Z_{p}e^{2}/\rho _{ioniz}$, where $%
\rho _{ioniz}$ is the impact parameter leading to ionization, given by Eq.(%
\ref{ionization impact parameter}). Substituting $\rho _{ioniz}$ from Eq.(%
\ref{ionization impact parameter}) into electron potential energy $%
Z_{p}e^{2}/\rho _{ioniz}$ gives that potential energy is larger than kinetic
energy if
\begin{equation}
v<v_{nl}.  \label{condition for electron capture}
\end{equation}%
Therefore, under the condition in Eq.(\ref{condition for electron capture}),
an electron can be effectively captured by the projectile after the
collision instead of leading to ionization. As a result, the ionization
cross section is small compared with the charge exchange cross section at
low projectile velocities. The assumption of the unperturbed electron wave
function results in grossly overestimated ionization cross sections as can
be seen in Fig.1.

The ionization cross sections are also difficult to measure at small
projectile energies, because careful separation between the large charge
exchange cross section and the small ionization cross section is necessary
for the correct measurement \cite{Shah}. Therefore, early measurements of
the ionization cross section at small velocities were not always accurate
\cite{Shvelko book, Shah}.

\subsubsection{Gillespie's treatment}

To account for the difference between the Born approximation results and the
experimental data for $v<v_{\max }$, Gillespie proposed to fit the cross
sections to the following function \cite{Gellipsie},%
\begin{equation}
\sigma ^{Gill}(v)=\exp \left[ -\lambda _{nl}\left( \frac{v_{0}\sqrt{Z_{p}}}{v%
}\right) ^{2}\right] \sigma _{mod}^{Bethe}(v).  \label{Gillespie}
\end{equation}%
Here, $\lambda _{nl}$ is a constant, which characterizes the ionized atom or
ion (for example, for the ground state of $H$, $\lambda _{nl}=0.76$), and $%
\sigma _{mod}^{Bethe}$ is the cross section in the Born approximation in the
form of Eq.(\ref{Bethe with Kim correction}). Gillespie's Eq.(\ref{Gillespie}%
) proved to fit very well existing experimental cross sections for hydrogen
atom ionization by $H^{+}$, $He^{+2}$, $Li^{+2}$,$Li^{+3}$, $C^{+4}$, $%
N^{+5} $, $N^{+4}$, $O^{+5}$ ions, and less well for $He$ and $H$ molecules
with the same ions \cite{Gellipsie}. Because $\sigma _{mod}^{Bethe}(v)$
becomes negative for $v<0.7$, Gillespie's Eq.(\ref{Gillespie}) can not be
applied to these low projectile velocities. In principle, the general fit $%
\sigma _{fit}^{BA}$ in Eq.(\ref{BA our fit paper}) can be used instead of $%
\sigma _{mod}^{Bethe}$ in Eq.(\ref{Bethe with Kim correction}). However,
because the two formulas differ considerably in the range of interest, $%
0.7<v<1$, the fitting coefficients $\lambda _{nl}$ have to be updated for
use with $\sigma _{fit}^{BA}$.

Although Gillespie's fit proved to be very useful, there are a number of
reasons to look for another fit. Gryzinski's Eq.(\ref{Gryzinski G(x)}) is
frequently used, because it requires only knowledge of one function for
calculations of cross sections, notwithstanding the fact that it
overestimates the cross sections at low energies.

\subsubsection{Bohr and Linhard's treatment}

For $v\lesssim v_{nl}$, a universal curve is expected if both the cross
sections and the square of impact velocity are divided by $Z_{p}$ \cite{Bohr
Linhard}. This scaling was established for the total electron loss cross
section $\sigma ^{el}$, which includes both the charge exchange cross
section $\sigma ^{ce}$ and the ionization cross section. Based on the
results of classical trajectory Monte Carlo (CTMC) calculations, Olson
proposed the following fit \cite{Olson},%
\begin{equation}
\sigma ^{el}(v,Z_{p})=Z_{p}A_{nl}\pi a_{0}^{2}f^{Olson}\left( \frac{v}{%
v_{0}\gamma _{nl}\sqrt{Z_{p}}}\right) ,  \label{Olson fit}
\end{equation}%
where $f(x)$ describes the scaled cross sections%
\begin{equation*}
f^{Olson}(x)=\frac{1}{x^{2}}\left[ 1-\exp \left( -x^{2}\right) \right] .
\end{equation*}%
Here, $\gamma _{nl}$ and $A_{nl}$ are constants, for example, $\gamma _{H}=%
\sqrt{5/4}=1.12$ and $A_{H}=16/3$ for atomic hydrogen, and $\gamma
_{He}=1.44 $ and $A_{he}=3.57$ for helium. The scaling in Eq.(\ref{Olson fit}%
) was also demonstrated analytically by Janev \cite{Janev}. For $v<<v_{0}%
\sqrt{Z_{p}}$, $\sigma ^{el}$ is dominated by charge exchange, $\sigma
^{ce}\approx \sigma ^{el}$, and Eq.(\ref{Olson fit}) gives a constant cross
section for charge exchange, $\sigma ^{ce}\approx \sigma ^{el}=16\pi
Z_{p}/3a_{0}^{2}$. For $v>>v_{0}\sqrt{Z_{p}}$, $\sigma ^{el}$ is dominated
by the ionization cross section, and $\sigma ^{ce}\approx \sigma
_{classical}^{high-energy}$ [Eq.(\ref{high energy Limit})]. Note that the
scaling in Eq.(\ref{Olson fit}) does not reproduce the logarithmic term in
the Bethe formula [Eq.(\ref{Bethe equation})] for $v>>v_{0}\sqrt{Z_{p}}$
because it is based on classical trajectory calculations. To make Eq. (\ref%
{Olson fit}) agree with Eq.(\ref{high energy Limit}), the coefficients $%
\gamma _{nl}$ should be proportional to $\sqrt{I_{nl}}$. For example, the
ionization potential for hydrogen is $I_{H}=13.6eV$, and for helium $%
I_{He}=24.6eV$. The ratio of $\gamma _{H}=1.12 $ to $\gamma _{He}=1.44$
differs from $\sqrt{I_{H}}/\sqrt{I_{He}}$ by only five percent, i.e., $%
\gamma _{H}/\sqrt{I_{H}}/\left( \gamma _{He}/\sqrt{I_{He}}\right) =1.05$.
Therefore, as was shown by Janev \cite{Janev}, the scaling in Eq.(\ref{Olson
fit}) can be rewritten\ in a form similar to Eq.(\ref{final classic result})
by normalizing the velocity to $v_{nl},$ Eq.(\ref{vnl on Inl}), i.e.,
\begin{equation}
\sigma ^{el}(v,I_{nl},Z_{p})=\pi a_{0}^{2}Z_{p}N_{nl}\,\,\frac{E_{0}^{2}}{%
I_{nl}^{2}}B_{nl}G^{el}\left( \frac{v}{v_{nl}\sqrt{Z_{p}}}\right) ,
\label{Janev Olson's generalization fit}
\end{equation}%
where
\begin{equation*}
G^{el}\left( x\right) =\frac{4}{3}f^{Olson}\left( x/\gamma _{H}\right) .
\end{equation*}%
Here, $N_{nl}$ is the number of electrons in the orbital $nl,$ and the $%
B_{nl}$ factors Eq.(\ref{Bnl}) are introduced to account for the
difference of the orbital electron velocity distribution functions
with the hydrogen-like EVDF function in Eq.(\ref{EVDF}). By
construction, Eq.(\ref{Janev Olson's generalization fit})
coincides with Eq.(\ref{high energy Limit}) in the limit
$v>>v_{nl}\sqrt{Z_{p}}$.

Because the scaling in Eq.(\ref{Olson fit}) is based on classical trajectory
calculations, it is valid only for intermediate velocities where the
underbarrier transitions allowed in the quantum mechanical calculations do
not contribute significantly (see Appendix B for details). Experimental data
\cite{Shah Li atoms, Janev} confirm the scaling in Eq.(\ref{Olson fit}) for $%
1.2<v/(v_{nl}\sqrt{Z_{p}})<3$, or equivalently, for the projectile energy in
the range $E=30-200\times Z_{p}I_{nl}/I_{H}$ in units of $keV/amu$.

A similar scaling to Eq.(\ref{Olson fit}) was derived in Ref.\cite{Duman}
based on quantum mechanical calculations making use of the quasi-classical
approach developed originally by Keldysh for multi-photon ionization of
atoms in a strong electromagnetic field. These calculations give scaling
similar to Eq.(\ref{Olson fit}), but with a different function $f(x)$ given
in \cite{Duman}. The quantum mechanical calculation results for the charge
exchange cross section in Ref. \cite{Duman} are a factor of $3$ larger than
Olson's cross section in Eq.(\ref{Olson fit}) for $v/(v_{0}\sqrt{Z_{p}})<0.2$%
.

Direct application of the scaling in Eq.(\ref{Janev Olson's generalization
fit}) for the ionization cross section instead of the total electron removal
cross section does not produce a single scaled function [see Fig.2 for
hydrogen and Fig. 4.(b) for helium]. Furthermore, the data are considerably
scattered near the maxima of the cross sections.
\begin{figure}[tbp]
\includegraphics{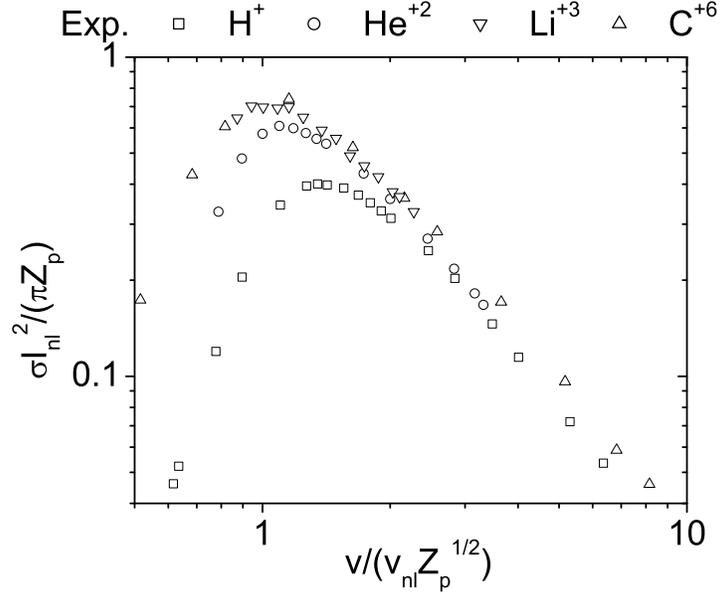} 
\caption{Ionization cross sections of hydrogen by fully stripped ions. The
scaled experimental data are from Fig.1. Note that the data do not merge
into a single curve. }
\label{Fig.2}
\end{figure}
A number of other semi-empirical models have been developed, which
use up to ten fitting parameters to describe the ionization cross
sections over the entire projectile energy range \cite{Daniel}.

\section{New fit formula for the ionization cross section}

\begin{figure}[tbp]
\includegraphics{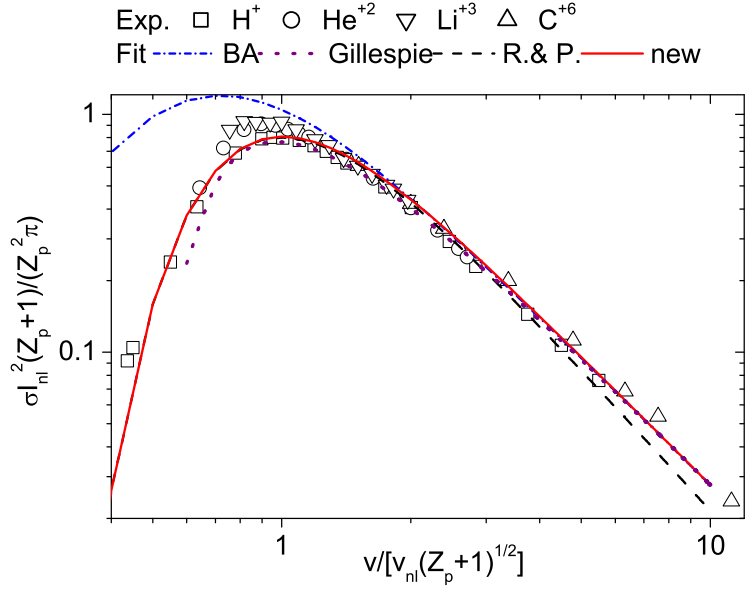}
\caption{Ionization cross sections of hydrogen by fully stripped ions
showing the scaled experimental data and the theoretical fits. BA denotes
the Born approximation [Eq.(\protect\ref{BA our fit paper})]. Gillespie
denotes Gillespie's fit according to Eq.(\protect\ref{Gillespie}). R.\&P.
symbolizes the fit proposed by Rost and Pattard \protect\cite{Rand P} in Eq.(%
\protect\ref{R&P scaling}). "New" denotes the new fit given by Eq.( \protect
\ref{New scaling fit formula}). }
\label{Fig.3}
\end{figure}

Analysis of the experimental data in Fig.\ref{Fig.1} shows that the maxima
of the experimentally measured cross sections occur at $\sqrt{Z_{p}+1},$ not
at $\sqrt{Z_{p}}$ as would be the case according to Olson's scaling in Eq.(%
\ref{Olson fit}). Therefore, it is natural to plot cross sections as a
function of the normalized velocity $v/(v_{nl}\sqrt{Z_{p}+1})$. Note that at
large velocities, according to Eq.(\ref{high energy Limit}) $\sigma \sim
Z_{p}^{2}/v^{2}$. Therefore, making use of the normalized velocity $v/(v_{nl}%
\sqrt{Z_{p}+1})$ requires normalization of the cross sections according to $%
\sigma /\left[ Z_{p}^{2}/(Z_{p}+1)\right] $. As a consequence, instead of
Eq.(\ref{Janev Olson's generalization fit}), we propose the following
scaling
\begin{equation}
\sigma ^{ion}(v,I_{nl},Z_{p})=\pi a_{0}^{2}\frac{Z_{p}^{2}}{(Z_{p}+1)}N_{nl}%
\frac{E_{0}^{2}}{I_{nl}^{2}}G^{new}\left( \frac{v}{v_{nl}\sqrt{Z_{p}+1}}%
\right) .  \label{New scaling}
\end{equation}%
Resulting plots of the scaled cross sections are shown in Fig.\ref{Fig.3}.
Comparing Fig.2 and Fig.3 one can clearly see that all of the experimental
data merge close to each other on the scaled plot based on Eq.(\ref{New
scaling}).

The resulting universal function can be fitted with various functions, but
the simplest fit was proposed by Rost and Pattard \cite{Rand P}. They showed
that if both the cross section and the projectile velocity are normalized to
the values of cross section and projectile velocity at the cross section
maximum, then the scaled cross section $\sigma /\sigma _{\max }$ is well
described by the fitting function
\begin{equation}
\sigma (v)=\sigma _{\max }\frac{\exp (-v_{\max }^{2}/v^{2}+1)}{v^{2}/v_{\max
}^{2}}.  \label{R&P scaling}
\end{equation}%
Here, $\sigma _{\max }$ is the maximum of the cross section, which occurs at
velocity $v_{\max }$. For the present study (the case of the ionization
cross section by the bare projectile), we predict that
\begin{eqnarray}
\sigma _{\max } &=&\pi a_{0}^{2}B_{nl}\frac{Z_{p}^{2}}{(Z_{p}+1)}\frac{%
E_{0}^{2}}{I_{nl}^{2}},  \label{Sigma maximum} \\
v_{\max } &=&v_{nl}\sqrt{Z_{p}+1},
\end{eqnarray}%
where the coefficients $B_{nl}$ depend only weakly on the projectile charge.
From Fig.3 one can estimate $B_{nl}=0.8$ for the ionization of hydrogen by
protons, while for ionization of hydrogen by bare nuclei of helium and
lithium, we find $B_{nl}=0.93$. As can be seen from Fig.3, the function in
Eq.(\ref{R&P scaling}) with $\sigma _{\max }$ and $v_{\max }$ defined in Eq.(%
\ref{Sigma maximum}) describes well the cross sections at small and
intermediate energies, but underestimates the cross section at high
energies. The reason is that the function in Eq.(\ref{R&P scaling}) does not
reproduce the logarithmic term in the Bethe formula in Eq.(\ref{Bethe
equation}). To improve the agreement with the experimental data and the
Bethe formula we propose a new scaling for the fitting function in Eq.(\ref%
{New scaling}) defined by%
\begin{equation}
G^{new}(x)=\frac{\exp (-1/x^{2})}{x^{2}}\left[ 1.26+0.283\ln \left(
2x^{2}+25\right) \right] .\;  \label{New scaling fit formula}
\end{equation}%
At large $x>>1$, Eq.(\ref{New scaling fit formula}) approaches the Bethe
formula in Eq.(\ref{G bethe}), and at small $x<1$, Eq.(\ref{New scaling fit
formula}) approaches the result in Eq.(\ref{R&P scaling}). The function $%
G^{new}(x)$ has a maximum at $x\simeq 1$, with $G^{new}(1)\simeq 0.86.$
Because $0.86$ is in between the maxima of the scaled cross section of
hydrogen by protons ($B_{nl}=0.8$) and the cross section for ionization of
hydrogen by bare nuclei of helium and lithium ($B_{nl}=0.93$), we did not
incorporate the coefficients $B_{nl}$ in Eq.(\ref{New scaling fit formula}).
This gives it a general form and introduces small errors of less than 8\%.

We have applied the new fit in Eqs.(\ref{New scaling}) and (\ref{New scaling
fit formula}) to the ionization cross section of helium, shown in Fig.4a.
The symbols in Fig.4a denote the experimental data for $H^{+}$, $%
He^{+2}$, $Li^{+3}$ \cite{Shah He 85, Shah He 89}, for $C^{+6}$ \cite{expC6}%
, for $I^{+Z_{p}}$ and $U^{+Z_{p}}$ \cite{expIandU}, and for
$Au^{+Z_{p}}$ \cite{expAu}, where $Z_{p}=10-40$. The solid curves
correspond to the continuum-distorted-wave-eikonal initial state
(CDW-EIS) theoretical calculation from Ref. \cite{He theory},
which is a generalization of the Born approximation. The CDW-EIS
theory accounts for the distortion of the electron wave function
by the projectile. From Fig.4a it is evident that the CDW-EIS
theory overestimates the cross section near the maximum, and
underestimates the cross section at small energies.

\begin{figure}[tbp]
\includegraphics[width=150mm]{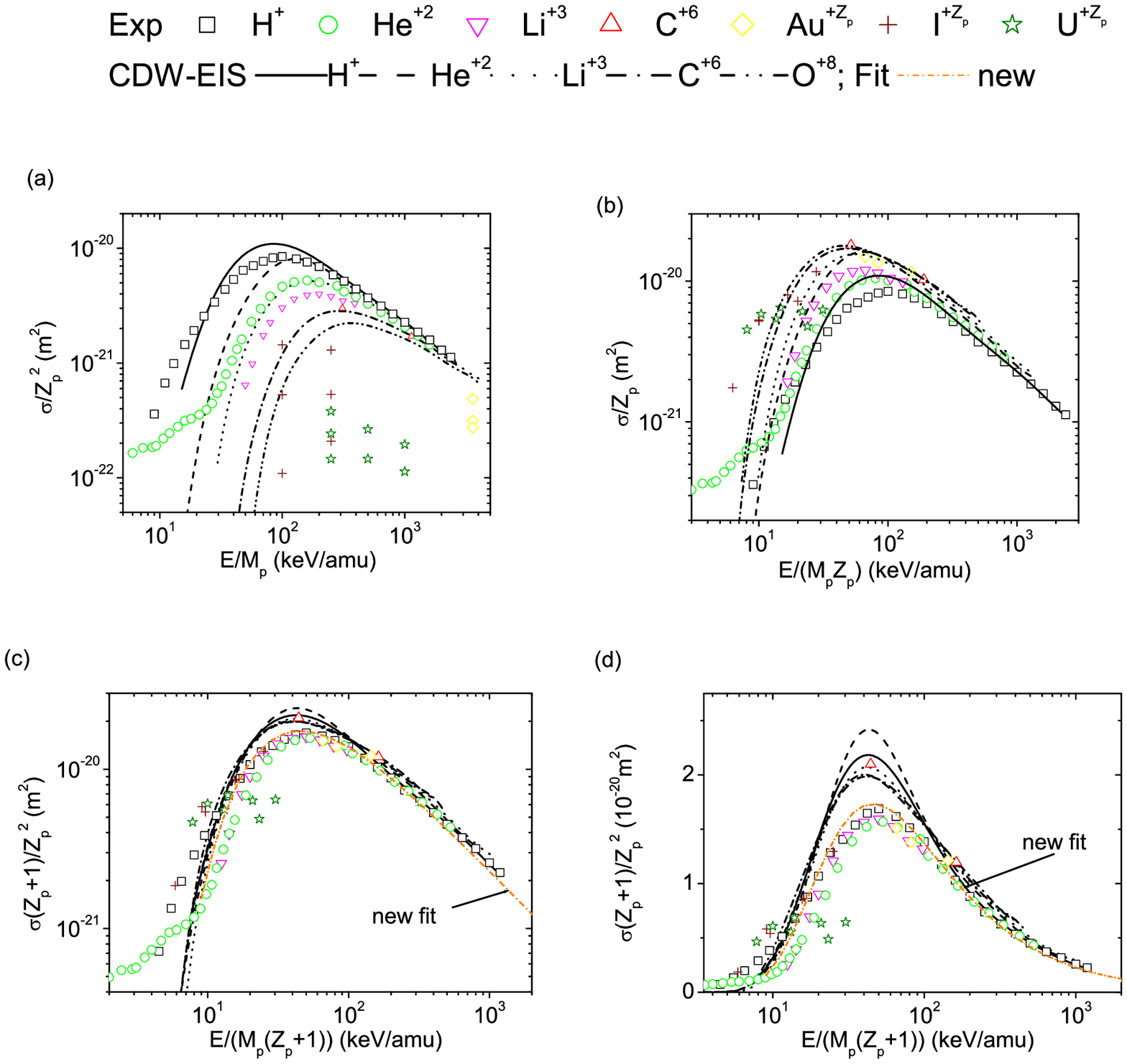}
\caption{Ionization cross sections of helium by various stripped
ions. The solid curves correspond to the CDW-EIS theoretical
calculation, and the symbols label the experimental data (see text
for details). Shown in the figures are: (a) the raw data; (b) the
scaled data from Fig.4a, making use of Eq.%
(\protect\ref{Janev Olson's generalization fit}); (c) the scaled
data making use of Eq.(\protect\ref{New scaling}); and (d) the
experimental data scaled using only Eq.(\protect\ref{New scaling})
together with the fit function. The notation "new fit" denotes
Eq.(\protect\ref{New scaling fit formula}).} \label{Fig. 4}
\end{figure}

Direct application of the scaling formula in Eq.(\ref{Janev Olson's
generalization fit}) to the ionization of helium does not produce similar
good results to the hydrogen case [see Fig. 4(b)]. But after applying the
new scaling in Eq.(\ref{New scaling}), all of the experimental and
theoretical results merge close together on the scaled plot, as is clearly
evident in Fig.4(c). Moreover, if we use the fit function of velocity
normalized to the orbital velocity $v_{nl}$ estimated from the ionization
potential of helium ($I_{He}=24.6$eV) making use of Eq.(\ref{vnl on Inl}),
the cross section is given by the same scaling as in Eq.(\ref{New scaling})
with the same function as in Eq.(\ref{New scaling fit formula}), as evident
from Fig.4(d). (The number of electrons in the helium atom is $N_{nl}=2$,
and therefore the scaled cross section is twice that of hydrogen.) From
Fig.4(d) it is clear that the new proposed fit in Eq.(\ref{New scaling})
using the function in Eq.(\ref{New scaling fit formula}) gives very good
results for both hydrogen and helium. Further verification of the new
scaling is difficult because reliable experimental data and numerical
simulations for a broad range of projectile velocities are absent for other
atoms. The discrepancy between the new fit and the helium data at very small
velocities is discussed in the next section.

Note that one experimental point in Fig.4 for $C^{+6}$ projectiles
is located far away from the fit. The error bar for this point is
about 30\% \cite{expC6}. This data may be inaccurate, as the
experimental point is higher than the predictions of CDW-EIS
theory, which overestimates the cross section near the maxima of
the cross sections for all other ions. The reason for the large
scatter in the uranium data on the scaled plot at small energies
is not clear, because the experimental data for all other
projectiles are located much closer to the fit line.

\section{Theoretical justification for the new fit formula for ionization
cross section}

In this section we discuss the theoretical foundations for the new fit to
the ionization cross section given by Eq.(\ref{New scaling}) and Eq.(\ref%
{New scaling fit formula}). We start with an analysis of high projectile
velocities.

\subsection{Behavior of cross sections at large projectile velocities $%
v>v_{nl}$}

\begin{figure}[tbp]
\includegraphics{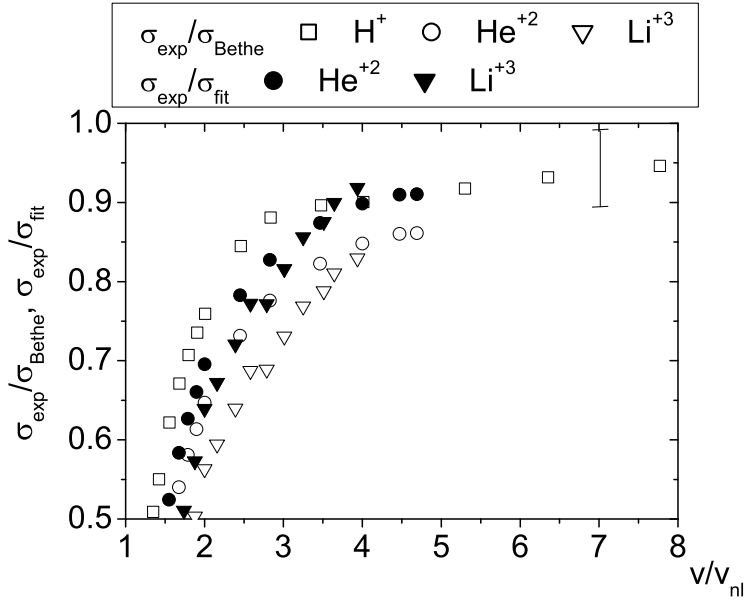}
\caption{Ratio of ionization cross sections of hydrogen by fully stripped
ions to the Bethe formula in Eq.(\protect\ref{Bethe equation}) and the fit
formula in Eq.(\protect\ref{fit high energy}) at high velocities. The
experimentally estimated uncertainty of 5.5\% \protect\cite{Shah Li atoms}
is shown by the error bar.}
\label{CompExpBethe}
\end{figure}

In the region of high projectile velocities the new fit predicts the
ionization cross section
\begin{equation}
\sigma _{fit}^{high-energy}(v)=4\pi a_{0}^{2}\,\,\frac{v_{0}^{4}}{v_{nl}^{2}}%
\frac{Z_{p}^{2}}{v^{2}}\left[ 0.566\ln \left( \frac{v}{v_{nl}\sqrt{%
(Z_{p}+1)/2}}\right) +1.26\right] ,  \label{fit high energy}
\end{equation}%
which differs from the Bethe formula in Eq.(\ref{Bethe equation}). [The
factor $\sqrt{(Z_{p}+1)/2}$ appears in the denominator under the logarithm
in the first term on the right hand side of Eq.(\ref{fit high energy}).] We
claim that incorporating this factor gives a better cross section estimate
than the Bethe formula. A comparison of the existing experimental data with
the Bethe formula in Eq.(\ref{Bethe equation}) and the fit formula in Eq.(%
\ref{fit high energy}) is shown in Fig.\ref{CompExpBethe}. The
experimentally estimated uncertainty of 5.5\% \cite{Shah Li atoms} is shown
by the error bar. The region of validity of the Born approximation and,
hence, the Bethe formula is \cite{Landau book, Bohr}%
\begin{equation}
v>\max (2Z_{p}v_{0},v_{nl}).  \label{BA validity}
\end{equation}%
The first condition in Eq.(\ref{BA validity}) assures that the projectile
potential is taken into account in the Born approximation; the second
condition allows use of the unperturbed atomic wave function. Unfortunately
the experimental data exists in the region in Eq.(\ref{BA validity}) only
for the ionization of hydrogen by protons. Figure \ref{CompExpBethe} shows
that the Bethe formula describes the experimental data for ionization of
hydrogen by protons within the error bar only for $v>6v_{0}$. Application of
the fit formula instead of the Bethe formula reduces discrepancy with the
data.

\ The applicability of the Born theory and the Bethe formula in Eq.(%
\ref{Bethe equation}) was studied experimentally in Refs. \cite{expC6,
expAu, q2scaling, Japan exp}. It was confirmed that the necessary condition
for the validity of the Bethe formula is given by the condition in Eq.(\ref%
{BA validity}). The failure of the Bethe formula for large $Z_{p}$ is
apparent from the experimental data for gold ions shown in Fig.4(a). The ion
velocity corresponds to $v=12v_{0}$ or $v=8.9v_{nl}$, whereas $%
Z_{p}=24,43,54$, and does not satisfy the condition in Eq.(\ref{BA validity}%
). As a result, the cross sections are much smaller than given by
the Bethe formula, as evident from Fig.4(a). (At large projectile
energies, all data merge to the Bethe formula, which corresponds
to a straight line in a logarithmic plot, similar to Fig.1.)


The applicability of the Bethe formula is limited by the validity of the
Born approximation. One of the easiest ways to correct it was suggested in
Ref.\cite{Duman}. Firstly, the Born approximation is considered, making use
of a classical trajectory for the projectile and a quantum mechanical
description in the Born approximation for the electron. In this
approximation, the probability of ionization or excitation is a function of
the impact parameter $\rho $. Here, for brevity, we shall consider only
ionization of the hydrogen atom. The projectile particle interacts with the
atomic electron with a potential energy $V(\mathbf{R,r}_{e})=-Z_{p}e^{2}/|%
\mathbf{R}-\mathbf{r}_{e}|$, where $\mathbf{R}(t)=\mathbf{\rho }+\mathbf{v}t$
is the classical trajectory of the projectile particle, and $\mathbf{r}_{e}$
describes the position of the electron relative to the nucleus of the atom.
For any impact parameter $\rho $, the probability of ionization is given by
the square of the transition amplitude
\begin{equation}
P_{BA}(\rho )=\frac{1}{\hbar ^{2}}\left\vert \int d\mathbf{r}_{e}\Psi
_{i}(r_{e})\left[ \int dte^{i\Delta Et/\hbar }V(\mathbf{R,r}_{e})\right]
\Psi _{f}^{\ast }(r_{e})\right\vert ^{2}.  \label{semiclassical}
\end{equation}%
Here, $\Delta E$ is the transferred energy in the transition, and $\Psi _{i}$
and $\Psi _{f}$ are the initial and final electron wave functions,
respectively. It can be shown that the calculations of ion-atom ionization
cross sections using the conventional Born approximation describing the
collision making use of momentum transfer (outlined in Appendix B) and the
semiclassical Born approximation making use of the assumption of the
straight line classical projectile trajectory [Eq.(\ref{semiclassical})] are
equivalent \cite{Bethe book}.

For large impact parameters $\rho >>a_{0}$, we can expand $V(\mathbf{R,r}%
_{e})$ in powers of $\mathbf{r}_{enl}/R$ according to%
\begin{equation}
V(\mathbf{R,r}_{e})=Z_{p}e^{2}\left( -\frac{1}{R}+\frac{\mathbf{R}\cdot
\mathbf{r}_{e}}{R^{3}}\right) .  \label{dipole appr}
\end{equation}%
The first term does not contribute to the matrix element in Eq.(\ref%
{semiclassical}) due to the orthogonality of the final and initial states.
Substituting Eq.(\ref{dipole appr}) into Eq.(\ref{semiclassical}) and
integrating in time yields \cite{Bethe book}
\begin{equation}
P_{BA}(\rho )=\left( \frac{2Z_{p}v_{0}}{\rho v}\right) ^{2}\left\vert \int d%
\mathbf{r}_{e}\Psi _{i}(r_{e})\Psi _{f}^{\ast }(r_{e})\left[ \frac{\omega
\rho }{v}x_{e}K_{1}\left( \frac{\omega \rho }{v}\right) +iz_{e}\frac{\omega
\rho }{v}K_{0}\left( \frac{\omega \rho }{v}\right) \right] \right\vert ^{2},
\label{PBA dipole}
\end{equation}%
where $\omega =\Delta E/\hbar $, and $K_{n}$ is the modified Bessel
function. Expanding the Bessel functions for small and large arguments, or
simply evaluating the integrand in Eq.(\ref{PBA dipole}) approximately, we
can approximate
\begin{equation}
\frac{\omega \rho }{v}K_{1}\left( \frac{\omega \rho }{v}\right) =\left\{
\begin{array}{c}
1,\;\frac{\omega \rho }{v}<1 \\
0,\;\frac{\omega \rho }{v}>1%
\end{array}%
\right\} ,
\end{equation}%
and neglect the second term on the right hand side in Eq.(\ref{PBA dipole}),
which is small compared with the first term. The probability of ionization
vanishes for $\rho >\rho _{\max }\simeq v/\omega =2a_{0}v/v_{0}$,
corresponding to the adiabatic limit. For $\rho >\rho _{\max }$, the
collision time $\rho _{\max }/v>a_{0}/v_{0}$ is much longer than the
electron circulation time around the nucleus, and the collision is
adiabatic. Consequently, the ionization probability is exponentially small
for $\rho >2a_{0}v/v_{0}$.

The square of electron dipole matrix element averaged over all possible
momenta of the ionized electron is \cite{Bethe}
\begin{equation}
\sum_{f}\int d\mathbf{r}_{e}\left\vert \Psi _{i}(r_{e})x_{e}\Psi _{f}^{\ast
}(r_{e})\right\vert ^{2}=0.283a_{0}^{2}.
\end{equation}%
Note that the sum over all final states including \emph{both} ionization and
excitation gives
\begin{equation}
\sum_{f}<0|x_{e}|f><f|x_{e}|0>=<0|x_{e}^{2}|0>=\frac{1}{3}%
<0|r_{e}^{2}|0>=a_{0}^{2}.
\end{equation}%
In this sum, $0.717$ corresponds to excitation, and $0.283$ corresponds to
ionization \cite{Bethe}.

For large impact parameters the momentum transfer to the electron is small
and we can neglect the electron kinetic energy of the ejected electron
compared with the ionization potential. As a result, $\Delta E\approx
I_{H}=E_{0}/2$ and $\omega =v_{0}/2a_{0}$ (in atomic units). Finally for $%
\rho >a_{0},$ the ionization probability is%
\begin{equation}
P_{BA}(\rho )\approx 0.283\left( \frac{2a_{0}v_{0}Z_{p}}{\rho v}\right)
^{2}\left\{
\begin{array}{c}
1,\;\rho <2a_{0}v/v_{0} \\
0,\;\rho >2a_{0}v/v_{0}%
\end{array}%
\right\} .  \label{PBA dipole final}
\end{equation}%
The ionization cross section is given by the integral
\begin{equation}
\sigma =2\pi \int_{0}^{\infty }P_{BA}(\rho )\rho d\rho .
\label{classical BA estimate}
\end{equation}%
For $\rho >a_{0}$, we can use Eq.(\ref{PBA dipole final}) to estimate $%
P_{BA}(\rho )$. For $\rho <a_{0}$, the dipole approximation in Eq.(\ref%
{dipole appr}) is not valid. To evaluate $P_{BA}(\rho )$ approximately for $%
\rho <a_{0}$, we can utilize the fact that $\int dte^{i\Delta Et/\hbar }V(%
\mathbf{R,r}_{e})$ is a weak function of $\rho $ for $\rho <a_{0}$, and
therefore $P_{BA}(\rho )\approx P_{BA}(a_{0})$. Substituting $P_{BA}(\rho
)\approx P_{BA}(a_{0})$ for $\rho <a_{0}$, and $P_{BA}(\rho )$ from Eq.(\ref%
{PBA dipole final}) for $\rho >a_{0}$, into Eq.(\ref{classical BA estimate})
gives
\begin{equation}
\sigma =8\pi a_{0}^{2}\cdot 0.283\frac{v_{0}^{2}Z_{p}^{2}}{v^{2}}\left[
\frac{1}{2}+\ln \left( \frac{2v}{v_{0}}\right) \right] ,
\label{BA cross section approx classical}
\end{equation}%
The first term in Eq.(\ref{BA cross section approx classical}) comes from
contributions of impact parameters $\rho <a_{0}$, and the second term
originates from contributions of large impact parameters $\rho >a_{0}$,
respectively. Comparison with the exact result in the Born approximation in
Eq.(\ref{Bethe equation}) shows that the contribution of impact parameters $%
\rho <a_{0}$ is underestimated, and $1/2$ should be replaced by $1.52$. The
above considerations are valid if the total probability of ionization and
excitation [$P_{BA}^{tot}(\rho )=\left( 2Z_{p}a_{0}v_{0}/\rho v\right) ^{2}$%
, for $\rho >a_{0}$] for the entire region of impact parameters is less than
unity, which requires $2Z_{p}v_{0}/v<1$. (Note that the total probability of
ionization and excitation is about $4$ times larger for ionization only.)

For $2Z_{p}v_{0}/v>1$, the total probability of the ionization and
excitation $P_{BA}^{tot}(\rho )$ calculated using the Born approximation is
more than unity, $P_{BA}^{tot}(\rho )>1$, for impact parameters $\rho <\rho
_{break}=2Z_{p}a_{0}v_{0}/v$, indicating the breakdown of the Born
approximation \cite{Duman}. Similar to the previous case, we can estimate
the ionization probability $P_{BA}(\rho )$ from Eq.(\ref{PBA dipole final})
for $\rho >\rho _{break}>a_{0}$ and assume $P_{BA}(\rho )\approx P_{BA}(\rho
_{break})=0.283$ for $\rho <\rho _{break}$. These considerations result in a
cross section estimate similar to the Bethe formula but with the logarithmic
term in the form $\ln (\rho _{\max }/\rho _{\min })=\ln
(v^{2}/v_{0}^{2}Z_{p})$, which gives
\begin{equation}
\sigma =8\pi a_{0}^{2}\cdot 0.283\frac{v_{0}^{2}Z_{p}^{2}}{v^{2}}\left[
\frac{1}{2}+\ln \left( \frac{v^{2}}{v_{0}^{2}Z_{p}}\right) \right] .
\label{sigma correted}
\end{equation}%
This calculation results in a smaller cross section than the Bethe formula
for $2Z_{p}v_{0}/v>1$. Note that in the above analysis we have used
unperturbed electron wave functions, which is valid only for $v>>v_{0}$.

While a number of smart semi-empirical ways to improve the first Born
approximation were developed \cite{Theod1, Theod2, Theod3}, the rigorous
approaches to improve the Bethe formula are based on the eikonal
approximation instead of the Born approximation \cite{McGuire}. The eikonal
approximation is justified if $ka_{nl}>1$, where $k$ is the projectile
particle wave vector $k=Mv/\hbar $, and the projectile kinetic energy is
large compared to the potential energy interaction with the target. For
heavy projectile particles with mass much larger than the electron mass,
these conditions are well satisfied. The ionization cross section in the
eikonal approximation is given by \cite{Landau book}%
\begin{equation}
\sigma =2\pi \int \frac{qdq}{k^{2}}|f(\mathbf{q})|^{2},
\label{sigma Glouber}
\end{equation}%
where $f(\mathbf{q})$ is the amplitude of ionization with momentum transfer $%
\mathbf{q}$
\begin{equation}
f(\mathbf{q})=\frac{k}{2\pi i}\int \rho d\rho <final|\exp \left( \frac{i\int
Vdz}{\hbar }-i\mathbf{q\cdot \rho }\right) |initial>.  \label{f Glouber}
\end{equation}%
The eikonal approximation in Eqs.(\ref{sigma Glouber}) and (\ref{f Glouber})
accounts approximately for all orders of the perturbation series, whereas
the Born approximation only make use of the first order. The calculations in
the eikonal approximation yield a formula similar to Eq.(\ref{sigma correted}%
) \cite{Matveev}. Note that the validity of the eikonal approximation in Eq.(%
\ref{f Glouber}) is limited to $v>>v_{0},$because the electron wave
functions $\Psi _{i}$ and $\Psi _{f}$ are assumed to be unperturbed atomic
functions. The influence of the projectile on the electron wave functions
has to be taken into account for $v\lesssim v_{0}$. This is typically
performed in the distorted wave approximation \cite{Shvelko book}.

Therefore, the correction to the Born approximation in Eq.(\ref{sigma
correted}) and the eikonal approximation give a formula similar to Eq.(\ref%
{fit high energy}) but with a factor $\alpha \sqrt{Z_{p}}$ ($\alpha $ is a
coefficient of order unity), instead of $\sqrt{(Z_{p}+1)/2}$. At large
velocities, both formulas give similar results.

\subsection{Behavior of cross sections at small projectile velocities $%
v<v_{nl}$}

If the projectile velocity is small compared with the orbital velocity, the
collision is adiabatic and the electron circulates many times around both
nuclei. The electronic energy states need to be determined in such a
quasimolecule as a function of the positions of both nuclei at a particular
time. In both the quantum mechanical and the classical approaches,
ionization is only possible if during the collision the initial and final
electronic terms cross at some instant. In classical mechanics this
corresponds to the so-called "$v/2$ mechanism". In a collisional system
comprised of two nuclei of equal charges (say ionization of hydrogen by a
proton), an electron which is exactly in between the two nuclei experiences
a very small electric field because the electric fields from both nuclei
exactly cancel for all times at this point. The electron can "ride" this
saddle point of the potential if its velocity is equal to one-half the
velocity of the projectile. The collision dynamics is illustrated in Fig.6.
\begin{figure}[tph]
\includegraphics[width=150mm]{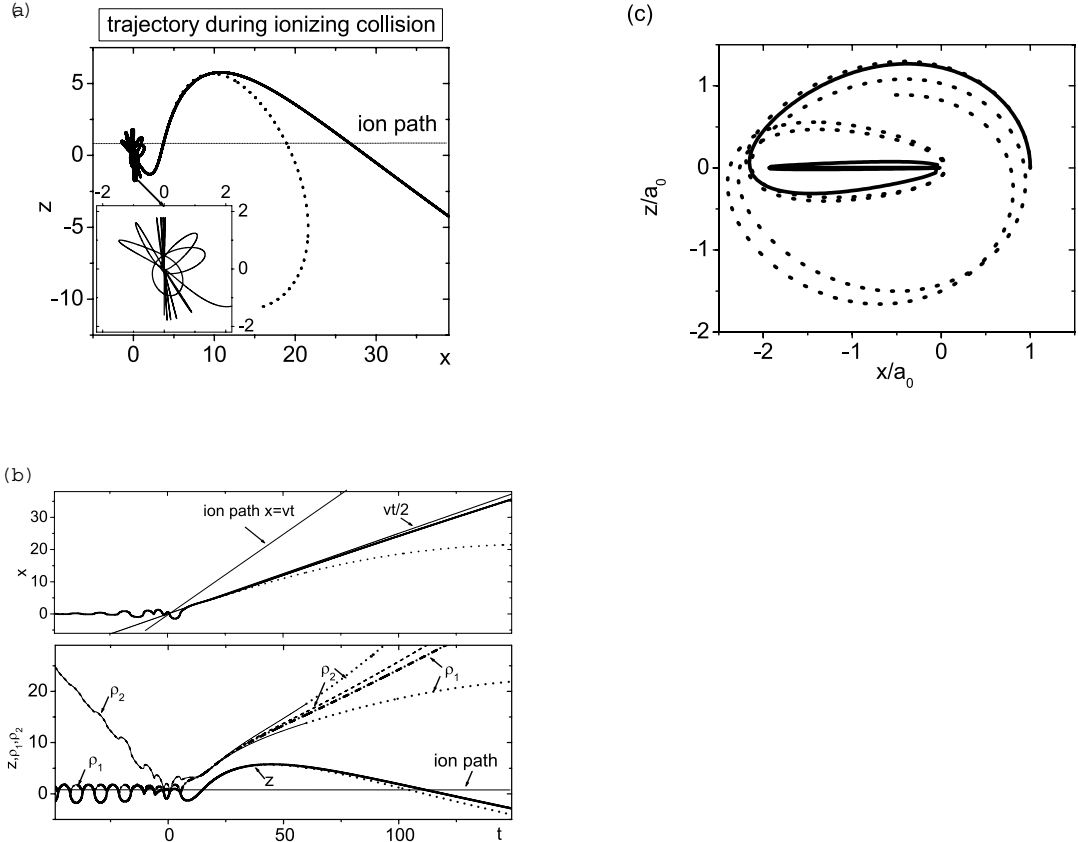}
\caption{The trajectory of a $v/2$ collision is shown in Figs.6(a) and (b).
The initial conditions correspond to a hydrogen atom with total energy $-1/2$%
, and at $t=-60$ $x=0=y$,$v_{x}=0=v_{y}$, $z=-1.606756$ (solid line) and $%
z=-1.606751$ (dotted line). The projectile moves along $z=1$ with velocity $%
1/2$. Atomic units are used: velocity is normalized to $v_{0}$; distance is
normalized to $a_{0}$; and time is normalized to $a_{0}/v_{0}$. Figure 6(b)
shows the position [$x(t),z(t)$] of the electron as a function of time, and
the distance between the electron and the first ($\protect\rho _{1}$) and
the second proton ($\protect\rho _{2})$ for the same conditions as in
Fig.6(a). The trajectory of a $S$-promotion is shown in Figs.6(c) for fixed
positive charges ($v\rightarrow 0$). The initial conditions correspond to an
internuclear separation $2a_{0}$ (in atomic units), initial position of the
electron $z=0,$ $x=1$; and initial velocity $v_{x}=0$, $v_{z}=1.155$ (solid
line), and $v_{z}=1.165$ (dotted line). }
\label{Fig.6}
\end{figure}

From Fig.6 one can see that the electron is stranded in between the protons
at $t=15a_{0}/v_{0}$ and its velocity projection on the x-axis is one-half
of the projectile velocity. A small variation of the initial condition from $%
z=-1.606756a_{0}$ (solid line) to $z=-1.606751a_{0}$ (dotted line)
completely changes the result of the collision. After the collision the
electron stays near the first nucleus and does not become ionized. As a
result, the probability of ionization is extremely small even though the
projectile velocity is not small (for the conditions in Fig.6, $v=1/2$ in
atomic units). The mechanism for ionization described above is also
so-called T-promotion in quantum mechanical descriptions \cite{Ovchnnikov}.

Another mechanism for ionization is attributed to the so-called S-promotion
mechanism \cite{Ovchnnikov}. It is associated with the special type of
trajectory of the electron in the field of two positive charges, shown in
Fig.6(c). Figure 6(c) shows that an electron with particular initial
conditions tends to spiral with a large number of turns enclosing a segment
of the straight line joining the nuclei Fig.6(c) \cite{Abramov}. Such a
trajectory is unstable - a small variation of initial conditions results in
a completely different trajectory as shown in Fig.6(c). Analysis of the
electron motion in the field of two positive charges, $Z_{T}$ and $Z_{P}$,
which are separated by a distance $R$ is best described in elliptical
coordinates
\begin{equation}
\xi =\frac{r_{p}+r_{T}}{R},\;\eta =\frac{r_{p}-r_{T}}{R},
\end{equation}%
where $r_{p}$ and $r_{T}$ are the distances from the electron to the
projectile and target nuclei, respectively. Making use of atomic units, the
classical trajectory in terms of the variables $\xi $ and $\eta $ can be
expressed as \cite{Abramov}
\begin{equation}
\frac{d\xi }{dt}=\frac{4(\xi ^{2}-1)P_{\xi }}{R^{2}(\xi ^{2}-\eta ^{2})},\;%
\frac{d\eta }{dt}=-\frac{4(\eta ^{2}-1)P_{\eta }}{R^{2}(\xi ^{2}-\eta ^{2})},
\label{dksi dt}
\end{equation}%
where the canonical momentums $P_{\xi }$ and $P_{\eta }$ are
\begin{eqnarray}
P_{\xi } &=&\left( -\frac{1}{2}R^{2}|E|+\frac{(Z_{P}+Z_{T})R\xi -\lambda }{%
\xi ^{2}-1}-\frac{P_{\phi }^{2}}{(\xi ^{2}-1)^{2}}\right) ^{1/2},\;
\label{Pksi} \\
P_{\eta } &=&\left( -\frac{1}{2}R^{2}|E|+\frac{(Z_{P}-Z_{T})R\eta +\lambda }{%
1-\eta ^{2}}-\frac{P_{\phi }^{2}}{(1-\eta ^{2})^{2}}\right) ^{1/2}.
\end{eqnarray}%
Here $E<0$ is the total energy of the electron, $P_{\phi }=\xi \eta d\phi
/dt $ is the rotational momentum around the straight line joining the
nuclei, and $\lambda $ is the integral of motion (for stationary nuclei)
\begin{equation}
\lambda =M^{2}-\frac{R^{2}}{4}\left( P_{\zeta }^{2}+\frac{P_{\phi }^{2}}{%
\zeta ^{2}}\right) +R(Z_{P}\cos \theta _{P}+Z_{T}\cos \theta _{T}).
\end{equation}%
Here, $\zeta $ is the closest distance from the electron to the straight
line joining the nuclei; $P_{\zeta }$ is the vector dot product of the
electron momentum with the $\zeta $-axis; $M^{2}=(\mathbf{r\times p)}^{2}$
is the total rotational momentum; and $\theta _{P}$ and $\theta _{T}$ are
the angles between $\mathbf{r}_{p}$ and $\mathbf{R,}$ and $\mathbf{r}_{T}$
and $-\mathbf{R}$, respectively. Moreover, $\mathbf{r}_{p}$ is the radius
vector from the projectile to the electron; $\mathbf{r}_{T}$ is the radius
vector from the target nucleus to the electron; and $\mathbf{R}$ is the
radius vector from the projectile to the target nucleus. The canonical
momentum $P_{\xi }$ in Eq.(\ref{Pksi}) tends to infinity if $\xi \rightarrow
1$, preventing the electron from approaching a segment of the straight line
joining the nuclei, $\xi =1$. In the special case
\begin{equation}
(Z_{P}+Z_{T})R=\lambda ,\;P_{\phi }=0,  \label{special condition}
\end{equation}%
the singularity vanishes at the point $\xi =1$ in Eq.(\ref{Pksi}). As a
result, for initial conditions satisfying the condition in Eq.(\ref{special
condition}), $P_{\xi }$ is finite for $\xi =1$. From Eq.(\ref{dksi dt}), $%
\xi $ approach unity exponentially with time -- the limiting electron
trajectory lies on the internuclear axis -- as shown in Fig.6(c), where the
initial conditions for the solid line correspond to the condition in Eq.(\ref%
{special condition}). A small departure from the condition in Eq.(\ref%
{special condition}) shown by the dotted line in Fig.6(c) prevents the
trajectory from approaching $\xi =1$. Thus the internuclear axis $\xi =1$,
represents the locus of points of unstable equilibria. In a quantum
mechanical treatment, such periodic unstable trajectories is responsible for
S-promotion of electron to the the continuum (ionization) when the nuclei
approach each other \cite{Ovchnnikov PRL}. The potential barrier in Eq.(\ref%
{Pksi}) increases when $R$ decreases. As a result, an electron near the top
of the barrier slows down and is then collected and promoted to the
continuum as the top of the barrier further rises. Due to the strong
instability of the locus, a numerical simulation of the corresponding
classical trajectory is extremely difficult. [We could not present the
classical analog of the ionization scenario for S-promotion, in contrast to
the T-promotion as shown in Fig. 6(a) and (b).]

The probability of ionization is greatly enhanced in quantum mechanics due
to tunnelling into classically forbidden regions of phase space. The cross
sections can be calculated using the quasiclassical method, where the
probability of transition is given by
\begin{equation}
P(\rho )=\exp \left( -\frac{2}{\hbar }\mathit{{Im}(S)}\right) ,
\label{quasiclassical}
\end{equation}%
where%
\begin{equation}
S(\rho ,\epsilon )=\sum_{n}\int_{c}pdR.
\end{equation}%
Here, $S(\rho ,\epsilon )$ is the classical action of the projectile ion,
and $p=\sqrt{2M(\epsilon -U(R,\rho )-E_{i})}$ is the projectile momentum,
generalized to classically forbidden regions of phase space where $p$ is
complex \cite{Landau book}. The integration contour in Eq.(\ref%
{quasiclassical}) is in the complex $R$ plane around the branch points ($%
R_{n}^{c}$) where the initial and final electronic terms cross [$%
E_{f}(R_{n}^{c})=E_{i}(R_{n}^{c})$]. Moreover, $n$ numerates different
branch points or channels of ionization for S and T-promotions. The
resulting cross section for hydrogen ionization by collision with a proton
is \cite{Ovchnnikov}%
\begin{equation}
\sigma _{adiabatic}(v)=\pi v\sum_{n}R_{n}^{2}e^{-2\Delta _{n}/v},
\label{Ovchnnikov}
\end{equation}%
where $n$ labels many different channels, and the coefficients
$\Delta _{n}$ and $R_{n}$ are of order unity in atomic units
($R_{n}$ is determined by the branch points $R_{cn})$. In the
range of projectile velocities $v=0.4-1$, we find that
Eq.(\ref{Ovchnnikov}) can be approximated to within 10\% accuracy
by only two exponents with $R_{1}=1.9$ , $\Delta _{1}=0.53$
(corresponding to S-promotion) and $R_{2}=6.7$, $\Delta _{2}=1.8$
(corresponding to T-promotion). Because $\Delta _{1}<<\Delta
_{2}$, primarily the S-promotion determines the ionization cross
section at small velocities ($v<0.5$), while both mechanisms
contribute to ionization for $v$ in the range $v=0.5-1$. Recent
experimental study and quantum mechanical calculations using
the continuum-distorted-wave eikonal-initial-state (CDW-EIS) model \cite%
{Shah 2000} show that a electron emission spectrum is dominated by
a well defined electron capture to continuum (S-promotion) peak
although existence of saddle-point electron emission (T-promotion)
is not confirmed.

The new fit predicts an extremely small cross section at very low velocity $%
\sigma _{fit}^{low-energy}(v)\sim \exp (-1/v^{2})$, whereas Eq.(\ref%
{Ovchnnikov}) gives $\sigma _{adiabatic}(v)\sim e^{-1.0/v}$. Therefore, the
numerical fit in Eq.(\ref{New scaling fit formula}) underestimates the cross
section for $v<0.5$, but gives a result close to the sum in Eq.(\ref%
{Ovchnnikov}) for $v$ in the range $v=0.5-1$. While the data for hydrogen at
very low projectile velocity is absent, and the fit agrees well for the
entire dataset in Fig.3, the disagreement is clearly seen when the fit is
compared with the experimental data for the ionization of He shown in
Fig.4(d). Adiabatic theory results are absent for helium, but the
experimental ionization cross section of He by protons can be described by
Eq.(\ref{Ovchnnikov}) with different coefficients $\Delta _{n}$ and $R_{n}$.
The behavior of the experimental ionization cross section of He by He$^{+2}$
is somewhat puzzling because of the very slow decrease of the cross section
for small projectile velocity.

In view of these observations, the applicability of the new fit is limited
to $v/[v_{nl}\sqrt{(Z_{p}+1)}]>0.5$. Note that for small projectile velocity
the ionization cross section is ten times smaller than the maximum of the
cross section, $\sigma _{\max }$, and the ionization cross section is
completely dominated by charge exchange, whose cross section is comparable
to $\sigma _{\max }$. Consequently both experimental measurements and
theoretical simulations are very difficult for very small projectile
velocity.

\section{Conclusions}

The new scaling in Eq.(\ref{New scaling}) for the ionization and stripping
cross sections of atoms and ions by fully stripped projectiles has been
proposed. The new scaling does not have any fitting parameters and describes
the shape of the cross section as a single function of the scaled projectile
velocity [Eq.(\ref{New scaling fit formula})]. Note that previous scaling
laws either used fitting parameters (\cite{Gillespie, Rand P}) or actually
did not match experiments in a wide range of projectile velocities \cite%
{Gerjuoy, Gryz}. The proposed scaling formula agrees well with theoretical
predictions in the limit of large projectile velocities. The new scaling has
been verified by comparison with available experimental data and theoretical
simulations for the ionization cross sections of hydrogen and helium by $%
H^{+},He^{+2},Li^{+3},C^{+6}$, and $O^{+8}$. The agreement between the new
proposed scaling and experimental data is very good. The difference between
the proposed fit and the experimental data is within 15\% accuracy, which is
similar to the estimated uncertainty in the measurements. The validity of
the fit is limited at very small velocities, where the ionization cross
section is very small, about one-tenth of the maximum cross section $\sigma
_{\max },$ and the ionization cross section is completely dominated by
charge exchange, whose cross section is comparable to $\sigma _{\max }$.
Finally, the fit is valid for scaled projectile velocity $v>0.5v_{nl}\sqrt{%
Z_{p}+1}$, where $v_{nl}=v_{0}\sqrt{2I_{nl}/E_{0}}$ is the orbital velocity
of the electron estimated from the ionization potential $I_{nl}$, where $%
E_{0}=27.2eV$ (twice the hydrogen ionization potential). Similarly, the fit
is valid for $E>12.5(Z_{p}+1)I_{nl}/E_{0}$ in units of $keV/amu$, where $E$
is the projectile kinetic energy per nucleon.

\textbf{Acknowledgments}

This research was supported by the U.S. Department of Energy Office of
Fusion Energy Sciences and the Division of High Energy Physics. It is a
pleasure to acknowledge the benefit of useful discussions with Scott
Armel-Funkhouser, Larry Grisham, Jun Hasegawa, Ed Lee, Dennis Mueller, David
R. Schultz, Ron Olson, Constantine E. Theodosiou, Lev D. Tsendin and Simon
Yu.

\appendix

\section{Classical cross section averaged over atomic electron velocity
directions}

Gerjuoy averaged the Rutherford cross section over all orientations of the
electron velocity $\mathbf{v}_{e}$ (for a fixed electron speed $v_{e})$ and
derived the differential cross section $d\sigma /d\Delta E(v_{e},v,\Delta E)$
for energy transfer $\Delta E$\ in the collision between a free electron and
the projectile \cite{Gerjuoy}. The total cross section is calculated by
integrating over values of energy transfer larger than the ionization
potential ($\Delta E>I_{nl}$\ ) and averaging over the electron velocity
distribution function (EVDF) $f\left( v_{e}\right) $. This gives%
\begin{equation}
\sigma (v,I_{nl},Z_{p})=Z_{p}^{2}\int_{0}^{\infty }\sigma
_{I_{nl}}(v,v_{e})\,f\left( v_{e}\right) \,\,dv_{e},  \label{Appendix A1}
\end{equation}%
where%
\begin{equation}
\sigma _{I_{nl}}(v,v_{e})=\int_{I_{nl}}^{\infty }\frac{d\sigma }{d\Delta E}%
(v,v_{e},\Delta E)d\Delta E,  \label{Appendix A2}
\end{equation}%
and $d\sigma /d\Delta E(v_{e},v,\Delta E)$\ is defined by \cite{Gerjuoy}

\begin{equation}
\frac{d\sigma }{d\Delta E}(v,v_{e},\Delta E)=\frac{\pi a_{0}^{2}}{4}\frac{%
E_{0}^{2}}{\Delta E^{3}}\frac{S(v,v_{e},\Delta E)}{v^{2}v_{e}},
\label{Gerjuoy differential}
\end{equation}%
where
\begin{equation*}
S(v,v_{e},\Delta E)=\left[
\begin{array}{c}
\left( v^{2}-v_{e}^{2}\right) \left( v_{e}^{2}-v^{2}-2\Delta E/m_{e}\right)
\left( v_{low}^{-1}-v_{up}^{-1}\right) + \\
2\left( v_{e}^{2}+v^{2}+\Delta E/m_{e}\right) \left( v_{up}-v_{low}\right)
-1/3\left( v_{up}^{3}-v_{low}^{3}\right)%
\end{array}%
\right] .
\end{equation*}%
Here, $v_{up}$ and $v_{low}$ are defined by
\begin{equation*}
v_{up}=v_{e}+v,
\end{equation*}%
\begin{equation*}
v_{low}=\max \left( \left\vert v_{e}-v\right\vert ,\sqrt{v_{e}^{2}-2\Delta
E/m_{e}}-v\right) .
\end{equation*}%
For very large projectile velocities $v>>v_{e}$, it follows that $S\approx
8v_{e}\left( 2v_{e}^{2}/3+\Delta E/m_{e}\right) $, and\ Eq.(\ref{Gerjuoy
differential}) yields%
\begin{equation}
\frac{d\sigma _{classical}^{high\quad energy}}{d\Delta E}(v,v_{e},\Delta
E)=2\pi a_{0}^{2}\frac{E_{0}^{2}}{\Delta E^{3}m_{e}v^{2}}\left( \frac{%
2m_{e}v_{e}^{2}}{3}+\Delta E\right) .  \label{Appendix A4}
\end{equation}%
Substitution of Eq.(\ref{Appendix A4}) into Eq.(\ref{Appendix A2}), and
subsequent substitution of Eq.(\ref{Appendix A2}) and the EVDF Eq.(\ref{EVDF}%
) into Eq.(\ref{Appendix A1}) give%
\begin{equation}
\sigma _{classical}^{high\quad energy}(v,I_{nl},Z_{p})=\frac{10}{3}\pi
Z_{p}^{2}a_{0}^{2}\frac{v_{0}^{2}E_{0}}{v^{2}I_{nl}}.  \label{Appendix A5}
\end{equation}%
In the general case with $v\sim v_{e}$, substituting the EVDF Eq.(\ref{EVDF}%
) into Eqs.(\ref{Appendix A2}) and (\ref{Appendix A1}) yields
\begin{equation}
\sigma _{classical}(v,I_{nl},Z_{p})=\pi a_{0}^{2}E_{0}^{2}\frac{Z_{p}^{2}}{%
I_{nl}^{2}}G_{classical}\left( \frac{v}{\sqrt{2I_{nl}/m_{e}}}\right) ,
\label{Appendix A6}
\end{equation}%
where
\begin{equation*}
G_{classical}(x)=\frac{1}{x^{2}}\int_{0}^{\infty }\int_{1/2}^{\infty }\frac{%
S(x\sqrt{2I_{nl}/m_{e}},v_{e},\Delta E)\,f\left( v_{e}\right) }{\Delta
E^{3}v_{e}}\,d\Delta Edv_{e}.
\end{equation*}%
The approximate formula for $G_{classical}(x)$ is given below in Eq.(\ref%
{G(x) classical}).

\section{The Born approximation}

Although the Born approximation is valid only for large projectile
velocities $v>>Z_{p}v_{0}$ \cite{Landau book}, the Born approximation does
give results close to the experimental data even outside its validity range
\cite{Bates}. Therefore, we have studied cross sections in the Born
approximation for the entire velocity range.

In the Born approximation, the ionization cross section for hydrogen atoms
by impact of fully stripped projectile atoms with charge $Z_{p}$ is given by
\cite{Shvelko book, Bethe, Bethe book},%
\begin{equation}
\sigma _{nl}^{BA}(v)=8\pi a_{0}^{2}Z_{p}^{2}\frac{v_{0}^{2}}{v^{2}}%
\int_{0}^{\infty }\frac{P_{I_{nl}}(q,v)}{q^{3}}dq,  \label{BA cross section}
\end{equation}%
where $P_{I_{nl}}(q,\widetilde{v})$ is the probability of ionization, and $%
qm_{e}v_{0}$ is the momentum transfer during the collision. We introducing
the velocity in atomic units $\widetilde{v}\equiv v/v_{0}$, and $%
P_{I_{nl}}(q,\widetilde{v})$is determined by \cite{Bethe}
\begin{equation}
P_{I_{nl}}(q,\widetilde{v})=\int_{0}^{\infty }\frac{dP(q,\kappa )}{d\kappa }%
\Theta \left( q-\frac{\frac{I_{nl}}{E_{0}}+\frac{1}{2}\kappa ^{2}}{%
\widetilde{v}}\right) d\kappa .  \label{P(q,k)}
\end{equation}%
Here, $\Theta (x)$ is the Heaviside function, and $dP(q,\kappa )/d\kappa $
is the differential probability of ejecting an electron with momentum $%
\kappa m_{e}v_{0}$ when the momentum transfer from the projectile is $%
qm_{e}v_{0}$,%
\begin{equation}
\frac{dP(q,\kappa )}{d\kappa }=\left\vert \left\langle \Psi _{\kappa }^{\ast
}(\mathbf{p})\Psi _{0}(\mathbf{p+q})\right\rangle \right\vert
^{2}=\left\vert \left\langle \Psi _{\kappa }^{\ast }(\mathbf{r})e^{i\mathbf{%
qr}}\Psi _{0}(\mathbf{r})\right\rangle \right\vert ^{2}.
\label{P(q,k) matrix element}
\end{equation}%
In Eq.(\ref{P(q,k) matrix element}), $\Psi _{\kappa }^{\ast }(\mathbf{p})$
and $\Psi _{\kappa }^{\ast }(\mathbf{r})$ are the wave functions of the
continuous spectrum (ionized electron) in momentum space and coordinate
space, respectively; $\Psi _{0}(\mathbf{p})$ and $\Psi _{0}(\mathbf{r})$ are
the wave functions of the ground state, and star ($^{\ast }$) denotes
complex conjugate. According to \cite{Bethe},%
\begin{equation}
\frac{dP(q,\kappa )}{d\kappa }=2^{8}\kappa q^{2}\frac{\left[ q^{2}+\frac{1}{3%
}(1+\kappa ^{2})\right] \exp \{-2/\kappa \arctan [2\kappa /(1+q^{2}-\kappa
^{2})]\}}{\left[ (q+\kappa )^{2}+1\right] ^{3}\left[ (q-\kappa )^{2}+1\right]
^{3}\left( 1-e^{-2\pi /\kappa }\right) }.  \label{dP(q,k)}
\end{equation}%
For $q>>1$, the function $dP(q,\kappa )/d\kappa $ has a sharp maximum at $%
\kappa =q$ \cite{Landau book}%
\begin{equation}
\frac{dP(q,\kappa )}{d\kappa }=\frac{8}{3\pi }\frac{1}{\left[ (q-k)^{2}+1%
\right] ^{3}},  \label{dP(q,k)/dk large q}
\end{equation}%
which simply means that the entire momentum $q$ is transferred to the
ionized electron momentum $\kappa $. At small $q<1$, $dP(q,\kappa )/d\kappa
\sim \kappa q^{2}$ and the width of the function $P(q,\kappa )$ as a
function of $\kappa $ is of order unity in atomic units.

\begin{figure}[tbp]
\includegraphics{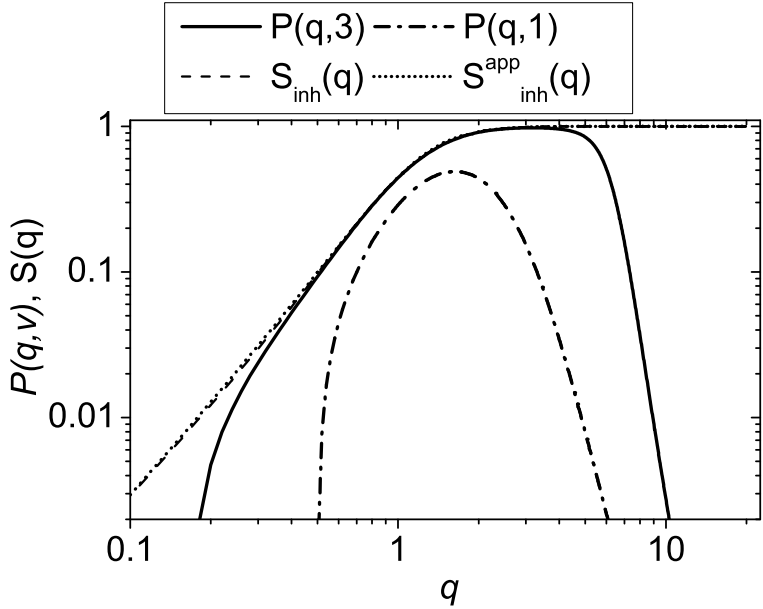}
\caption{Total ionization transition strength for atomic hydrogen as a
function of transferred momentum $q$. The exact function $P(q,v)$ [Eq.(
\protect\ref{P(q,k)})] for $\widetilde{v}=1$ and $\widetilde{v}=3$ is
compared with the approximate function $S_{inh}(q)$ [Eq.(\protect\ref{Sinh}%
)] (which is independent of $v$) and the fit $S_{inh}^{app}(q)$ in Eq.(%
\protect\ref{Sinh_appr}).}
\label{fig:Sinh}
\end{figure}

For large projectile velocity $v>>v_{0}$, considerable simplification can be
made by neglecting the electron kinetic energy $\frac{1}{2}\kappa ^{2}$ in
the argument of the Heaviside function in Eq.(\ref{P(q,k)}). The
approximation%
\begin{equation}
\Theta \left( q-\frac{\frac{I_{nl}}{E_{0}}+\frac{1}{2}\kappa ^{2}}{\frac{v}{%
v_{0}}}\right) \rightarrow \Theta \left( q-\frac{\frac{I_{nl}}{E_{0}}}{\frac{%
v}{v_{0}}}\right)  \label{Thetha substitution}
\end{equation}%
is referred to as the close-coupling approximation. In this case, $P(q,v)$
can be characterized by a function of one argument, $S_{inh}(q)$, with%
\begin{equation}
P_{I_{nl}}(q,\widetilde{v})=S_{inh}(q)\Theta \left( q-\frac{v_{0}I_{nl}}{%
vE_{0}}\right) ,  \label{P approximate}
\end{equation}%
where%
\begin{equation}
S_{inh}(q)=\int_{0}^{\infty }\frac{dP(q,\kappa )}{d\kappa }d\kappa .
\label{Sinh}
\end{equation}%
The function $S_{inh}(q)$ is refereed to as the total ionization transition
strength \cite{Gillespie}. Substituting Eq.(\ref{Thetha substitution})
results in artificial, additional contributions to the integral in Eq.(\ref%
{P(q,k)}) for $\kappa >\kappa _{add}=\sqrt{2(qv/v_{0}-I_{nl}/E_{0})}$. For
large projectile velocities $v>>v_{0}$ and $q>>1$, $\kappa _{add}\simeq
\sqrt{2qv/v_{0}}.$ The function $dP(q,\kappa )/d\kappa $ has a sharp maximum
at $\kappa =q$ [see Eq.(\ref{dP(q,k)/dk large q})]. Therefore the artificial
additions for $\kappa >\kappa _{add}$\ do not contribute to the integral if $%
\kappa _{add}>q$, which corresponds to $q<2v$, and the substitution in Eq.(%
\ref{Thetha substitution}) is valid$.$ In the opposite case of large
projectile velocities $v>>v_{0}$ but small $q$, it follows that $q\sim
v_{0}I_{nl}/(vE_{0})<<1$ , for the range of $q$ $\kappa _{add}\sim 1,$ and
the function $dP(q,\kappa )/d\kappa $ decreases rapidly for $\kappa >1$.
Therefore, the artificial additions for $\kappa >\kappa _{add}$\ do not
contribute to the integral if $\kappa _{add}>1$. Hence, the substitution in
Eq.(\ref{Thetha substitution}) is valid for $v>>v_{0}$. Figure 7 shows plots
of $P_{I_{nl}}(q,\widetilde{v})$ [Eq.(\ref{P(q,k)})] and $S_{inh}(q)$ [Eq.(%
\ref{Sinh})] for $\widetilde{v}=1$ and $\widetilde{v}=3$. At small
projectile velocities $v<v_{0}$, the substitution in Eq.(\ref{Thetha
substitution}) produces a considerable error [see Fig.7]. For repetitive
calculations, the function $S_{inh}(q)$ in Eq.(\ref{Sinh}) can be
approximated to within 3\% accuracy by%
\begin{equation}
S_{inh}^{app}(q)=\left[
\begin{array}{cc}
\frac{0.545q^{2}}{(q-0.9)^{2}+1.21} & q<2 \\
\tanh (0.8q) & q\geq 2%
\end{array}%
\right] .  \label{Sinh_appr}
\end{equation}%
The functions $S_{inh}(q)$ [Eq.(\ref{Sinh})] and $S_{inh}^{app}(q)$ [Eq.(\ref%
{Sinh_appr})] are shown in Fig. \ref{fig:Sinh}.

Having estimated the function $P_{I_{nl}}(q,\widetilde{v})$, the total cross
section can be evaluated analytically for large $v>>v_{0}$. The region of
small $q$ contributes significantly to the cross section [see Eq.(\ref{BA
cross section})]. Therefore, we split the integration in Eq.(\ref{BA cross
section}) into the two regions $q<q_{up}$ and $q>q_{up}$, where $q_{up}=1/2$%
. In the first region $q<q_{up}$, it follows that $P_{I_{nl}}(q,v)\approx
S_{inh}^{app}(q)\approx 0.283q^{2}$, and the integration in Eq.(\ref{BA
cross section}) gives
\begin{equation}
\int_{0}^{q_{up}}dq\frac{P_{I_{nl}}(q,v)}{q^{3}}\approx \int_{q_{\min
}}^{q_{up}}dq\frac{0.283}{q}=0.283\ln (q_{up}/q_{\min }),
\label{approx Born small q}
\end{equation}%
where$\ q_{\min }=v_{0}I_{nl}/vE_{0}$. In the second region, only the range
of $q_{up}<q<2$ contributes to the integral, because at large $q>>1$, $%
P_{I_{nl}}(q,v)/q^{3}\approx 1/q^{3}$ and the contribution to the integral
for large $q$ quickly decreases to zero. At very large $q>2v$, $%
P_{I_{nl}}(q,v)$ became smaller than unity, but this region does not
contribute to the integral and can be neglected. As a result, the integral $%
\int_{q_{up}}^{\infty }dqP_{I_{nl}}(q,v)/q^{3}$ does not depend on $v$ (for
the large $v$ under consideration). The integration from $q_{up}$ to
infinity gives $\int_{q_{up}}^{\infty }dqP_{I_{nl}}(q,v)/q^{3}\approx 0.666$%
, and finally the result is similar to the Bethe formula in Eq.(\ref{Bethe
equation}) with%
\begin{equation}
\sigma ^{Bethe}(\widetilde{v})=8\pi a_{0}^{2}\,\,\frac{Z_{p}^{2}}{\widetilde{%
v}^{2}}\left[ 0.283\ln \left( \widetilde{v}\right) +0.666\right] .
\label{Bethe appendix}
\end{equation}%
The small differences from the Bethe formula are due to utilization of the
close coupled approximation in Eq.(\ref{approx Born small q}), which
overestimates $P_{I_{nl}}(q,v)$ at small $q$ [see Fig.\ref{fig:Sinh}].

Comparison with the exact calculation (Fig.1) shows that the Bethe
asymptotic result is close to the exact calculation in Eq.(\ref{BA cross
section}) for $\widetilde{v}>2.$ To extend the Bethe formula to lower
velocities, the second-order correction in the parameter $v_{0}/v$ has been
calculated in \cite{Kim}, yielding the cross section in the form%
\begin{equation}
\sigma _{mod}^{Bethe}(\widetilde{v})=4\pi a_{0}^{2}\,\,\frac{Z_{p}^{2}}{%
\widetilde{v}^{2}}\left[ 0.57\ln \left( \widetilde{v}\right) +1.26-0.66\frac{%
1}{\widetilde{v}^{2}}\right] ,  \label{Bethe + Kimequation}
\end{equation}%
where $\widetilde{v}=v/v_{0}$. Equation (\ref{Bethe + Kimequation}) agrees
with the exact calculation in Eq.(\ref{BA cross section}) to within 10\% for
$\widetilde{v}>1.1$. We have developed the following fit for the cross
section in the Born approximation,%
\begin{equation}
\sigma _{fit}^{BA}(\widetilde{v})=4\pi a_{0}^{2}\,\,\frac{Z_{p}^{2}}{%
\widetilde{v}^{2}}\left[ 0.283\ln \left( \widetilde{v}^{2}+1\right) +1.26%
\right] \exp \left[ -\frac{1.95}{\widetilde{v}(1+1.2\widetilde{v}^{2})}%
\right] ,  \label{BA our fit}
\end{equation}%
which agrees with the exact calculation in Eq.(\ref{BA cross section}) to
within 2\% for $\widetilde{v}>1$, and to within 20\% for $0.2<\widetilde{v}%
<1.$

The previous analysis was performed for the hydrogen atom. In the case of
hydrogen-like electron orbitals, the similarity principle can be used. The
quantity $dP(q,\kappa )/d\kappa $ is identical for different electron
orbitals if $q,\kappa $ are scaled with the factor $1/Z_{T}=v_{0}/v_{nl}$
\cite{Landau book}. Therefore, $P_{nl}(q,v)=P_{H}(qv_{0}/v_{nl},v/v_{nl})$,
where $H$ denotes hydrogen atom, and
\begin{equation}
\sigma _{fit}^{BA}\left( \widetilde{v}=\frac{v}{v_{nl}}\right) =4\pi
a_{0}^{2}\,\frac{v_{0}^{4}}{v_{nl}^{4}}\,\frac{Z_{p}^{2}}{\widetilde{v}^{2}}%
\left[ 0.283\ln \left( \widetilde{v}^{2}+1\right) +1.26\right] \exp \left[ -%
\frac{1.95}{\widetilde{v}(1+1.2\widetilde{v}^{2})}\right] ,
\label{BA our fit for all}
\end{equation}%
where
\begin{equation}
\widetilde{v}=\frac{v}{v_{nl}}=\frac{v}{\sqrt{2I_{nl}/m_{e}}}.
\label{v scaled}
\end{equation}%
As we have noted for helium, most scalings can be used even for
non-hydrogen-like electron orbitals, provided the relationship in Eq.(\ref{v
scaled}) is used.

\subsection{Comparison between the quantum mechanical and classical
trajectory calculations for $v>>v_{nl}$}

\begin{figure}[tbp]
\includegraphics{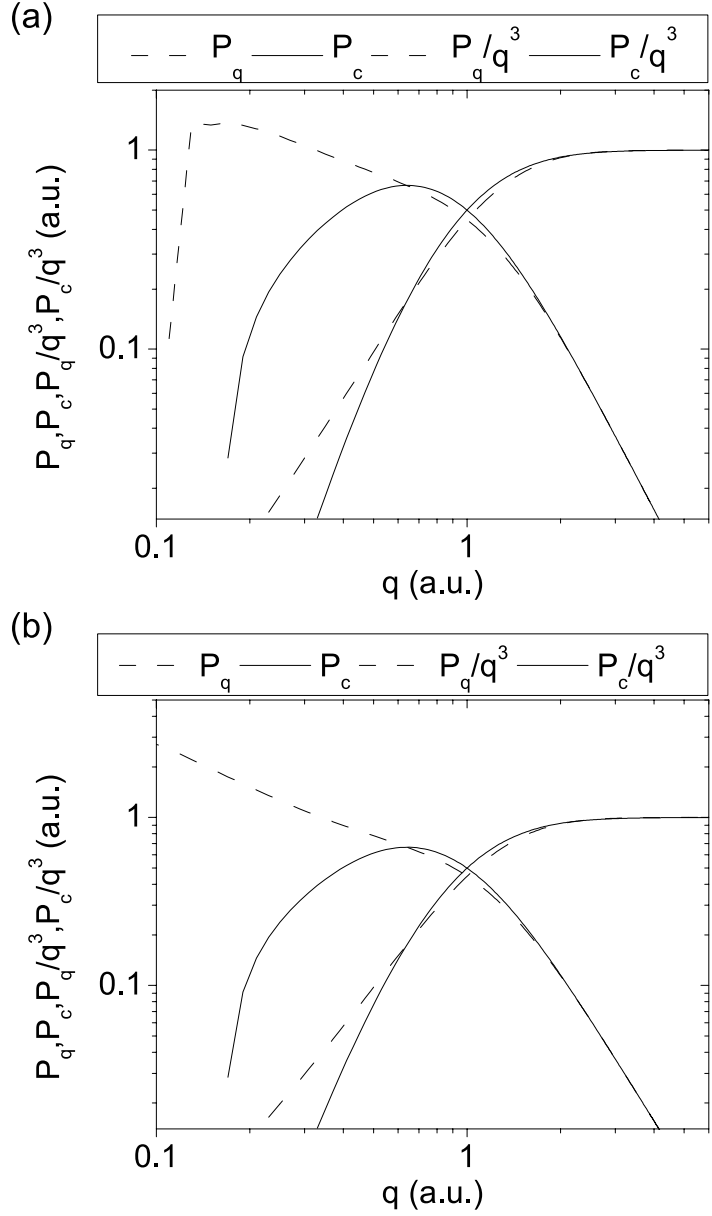}
\caption{Probability of ionization of atomic hydrogen as a function of
transferred momentum; $P_{c}(q)$ is given by classical mechanics [Eq.(%
\protect\ref{P classical x})], and $P_{q}(q,v)$ is given by quantum
mechanics [Eq.( \protect\ref{P(q,k)})]. The plots correspond to (a) $%
\widetilde{v}=5$ and (b) $\widetilde{v}=15$.}
\label{Figure 8}
\end{figure}
We have previously noted that the classical trajectory calculation
underestimates the ionization cross section at large velocities $v>>v_{nl}$.
To compare the ionization cross section calculated in the classical
trajectory and Born approximations, we present both cross sections in the
form of Eq.(\ref{BA cross section}). In the limit $v>>v_{nl}$, the momentum
transferred to the electron during a collision with impact parameter $\rho $
is given by Eq.(\ref{velocity kick}), i.e.,
\begin{equation}
q_{x}(\rho )\equiv m_{e}\Delta v_{x}(\rho )=\frac{2e^{2}Z_{p}}{v\rho },
\label{q transfer}
\end{equation}%
where $x-$axis is chosen in the direction perpendicular to the projectile
ion trajectory along the momentum transfer. Because $v>>v_{nl}$, the
electron velocity is neglected in Eq.(\ref{q transfer}). In classical
mechanics, ionization occurs if the energy transfer to the electron is more
than the ionization potential, $[(m_{e}\mathbf{v}_{e}+\mathbf{q}%
)^{2}-m_{e}^{2}v^{2}]/2m_{e}>I_{nl}$.

A small momentum transfer to the electron along the projectile trajectory $%
q_{z}(\rho )$ can be determined making use of the energy conservation. Due
to conservation of the momentum, the momentum transferred from the
projectile particle is $-q_{z}(\rho )$. The projectile energy change is $[(M%
\mathbf{v}-\mathbf{q})^{2}-M^{2}v^{2}]/2M=-vq_{z}$. Conservation of energy
gives%
\begin{equation}
vq_{z}\equiv \frac{1}{2m_{e}}[(m_{e}\mathbf{v}_{e}+\mathbf{q}%
)^{2}-m_{e}^{2}v_{e}^{2}].  \label{qz(qperp)}
\end{equation}%
In the limit $v>>v_{e}$, it follows that $q_{z}<<q_{x},$ and consequently
the total transferred momentum to the electron is $q=\sqrt{q_{\mathbf{x}%
}^{2}+q_{z}^{2}}\simeq q_{x}$. The momentum of the ejected electron can be
determined from the energy conservation relation
\begin{equation}
\kappa ^{2}/2m_{e}=[(m_{e}\mathbf{v}_{e}+\mathbf{q}%
)^{2}-m_{e}^{2}v_{e}^{2}]/2m_{e}-I_{nl}.  \label{k (qperp)}
\end{equation}%
In classical mechanics, the ionization probability of the ejected electron
with momentum $\kappa $ in a collision with total momentum transfer $q$ is
given by the integral over the electron distribution function,%
\begin{equation}
\frac{dP_{c}(q,\kappa )}{d\kappa }=\frac{\kappa }{m_{e}}\int f(\mathbf{v}_{e}%
\mathbf{)dv}_{e}\delta \left( \frac{\kappa ^{2}}{2m_{e}}-q_{x}v_{x}-\frac{%
q^{2}}{2m_{e}}-I_{nl}\right) .  \label{P(q,k) classical}
\end{equation}%
Introducing the one-dimensional electron distribution function
\begin{equation}
f_{x}(v_{ex})=\int f(\mathbf{v}_{e}\mathbf{)}dv_{y}dv_{z}\mathbf{,}
\label{EVDF x definition}
\end{equation}%
and substituting $q\simeq q_{x}$, Eq.(\ref{P(q,k) classical}) simplifies to
become
\begin{equation}
\frac{dP_{c}(q,\kappa )}{d\kappa }=\frac{\kappa }{qm_{e}}f_{x}\left( \frac{%
\kappa ^{2}-q^{2}-2m_{e}I_{nl}}{2qm_{e}}\right) .  \label{P(q,k) classical 2}
\end{equation}%
For hydrogen-like electron orbitals given by Eq.(\ref{EVDF}), $f_{x}(v_{ex})$
can be readily calculated to be
\begin{equation}
\,f_{x}\left( v_{ex}\right) =\frac{8}{3\pi }\frac{v_{nl}^{5}}{\left[
v_{ex}^{2}+v_{nl}^{2}\right] ^{3}}.  \label{EVDF x H}
\end{equation}%
Substituting the hydrogen-like electron distribution function Eq.(\ref{EVDF
x H}) into Eq.(\ref{P classical}) gives in atomic units
\begin{equation}
\frac{dP_{c}(q,\kappa )}{d\kappa }=\frac{16\kappa }{3\pi }\frac{%
(2qm_{e})^{5}v_{nl}^{5}}{\left[ \left( \kappa ^{2}-q^{2}-2m_{e}I_{nl}\right)
^{2}+(2qm_{e}v_{nl})^{2}\right] ^{3}}.  \label{dP(q,k)/dk large q classical}
\end{equation}%
Let us compare Eq.(\ref{dP(q,k)/dk large q classical}) with the quantum
mechanical result Eq.(\ref{dP(q,k)/dk large q}). In the limit $q>>1$, $%
\kappa \approx q$ and the two functions are equivalent. Both functions $%
dP(q,\kappa )/d\kappa $ have a maximum at $\kappa =q$, and the width of the
maximum is of order $1,$ which simply means that the entire momentum $q$ is
transferred to the ionized electron momentum $\kappa $.

Moreover it is possible to prove that the classical mechanical $%
dP_{c}(q,\kappa )/d\kappa $ is equivalent to the quantum mechanical function
$dP_{q}(q,\kappa )/d\kappa $ for any $s-$electron orbital (spherically
symmetrical wave function). Indeed, for large $k>>1$, the ejected electron
can be described as a sum over plane waves $\Psi _{\kappa }^{\ast }(\mathbf{r%
})\approx e^{i\mathbf{kr}}$, and substituting $\Psi _{\kappa }^{\ast }(%
\mathbf{r})$ into Eq.(\ref{P(q,k) matrix element}) gives
\begin{equation}
\frac{dP_{q}(q,\kappa )}{d\kappa }=\frac{1}{(2\pi \hbar )^{3}}\int
\left\vert \left\langle e^{i(\mathbf{q-k)r/}\hbar }\Psi _{0}(\mathbf{r}%
)\right\rangle \right\vert ^{2}k^{2}do_{\mathbf{k}}=\frac{1}{m_{e}^{3}}\int
f\left( \frac{\mathbf{q-k}}{m_{e}}\right) k^{2}do_{\mathbf{k}},
\label{P(q,k) classical 3}
\end{equation}%
where integral over $do_{\mathbf{k}}=2\pi \sin \vartheta d\vartheta $
designates averaging over all directions of the $\mathbf{k}$-vector, $%
\vartheta $ is the angle between $\mathbf{q}$ and $\mathbf{k}$, and $f\left(
v_{e}\right) $ is the electron distribution function in velocity space. Note
that $|\mathbf{q-k|}^{2}\mathbf{=q}^{2}\mathbf{+k}^{2}-2\mathbf{q\cdot k}%
=(q-k)^{2}+4qk\sin \vartheta /2^{2}$. In the limit $q>>1$, $k\approx q$ and
only small $\vartheta $ contribute to the integral in Eq.(\ref{P(q,k)
classical 3}). Therefore, averaging over all directions of the $\mathbf{k}$%
-vector gives
\begin{equation}
\frac{1}{m_{e}^{2}}\int f\left( \frac{\mathbf{q-k}}{m_{e}}\right) k^{2}do_{%
\mathbf{k}},=\frac{1}{m_{e}^{2}}\int f\left( \frac{\sqrt{(q-k)^{2}+qk%
\vartheta ^{2}}}{m_{e}}\right) \,2\pi k^{2}\vartheta d\vartheta .
\label{f averaged}
\end{equation}%
Introducing $v_{\perp }=k\vartheta /m_{e}$, the integral in Eq.(\ref{f
averaged}) takes form%
\begin{equation}
\int f\left( \sqrt{\left( \frac{q-k}{m_{e}}\right) ^{2}+v_{\perp }^{2}}%
\right) d^{2}v_{\perp }=f_{x}\left( \frac{q-k}{m_{e}}\right) ,  \label{f 1D}
\end{equation}%
where $f_{x}$ is the one-dimensional electron velocity distribution
function. Substituting Eqs.(\ref{f 1D}) and (\ref{f averaged}) into Eq.(\ref%
{P(q,k) classical 3}) yields
\begin{equation}
\frac{dP_{q}(q,\kappa )}{d\kappa }=\frac{1}{m_{e}}f_{x}\left( \frac{q-k}{%
m_{e}}\right) .  \label{Pq final}
\end{equation}%
Note that in the limit $q>>m_{e}v_{nl}$, it follows that $\kappa \approx q$,
and Eq.(\ref{P(q,k) classical 2}) becomes
\begin{equation}
\frac{dP_{c}(q,\kappa )}{d\kappa }=\frac{1}{m_{e}}f_{x}\left( \frac{q-k}{%
m_{e}}\right) .  \label{Pc final}
\end{equation}%
Finally, comparing Eqs.(\ref{Pq final}) and (\ref{Pc final}) we arrive at
the equivalence of functions $dP(q,\kappa )/d\kappa $ in quantum mechanics
and classical mechanics in the limit $q>>m_{e}v_{nl}$.

The situation is completely different for small $q<<m_{e}v_{nl}$. From Eq.(%
\ref{dP(q,k)/dk large q classical}) it follows that $dP_{c}(q,\kappa
)/d\kappa \sim \kappa q^{5}$, and $dP_{c}(q,\kappa )/d\kappa $ is much
smaller than $dP_{q}(q,\kappa )/d\kappa \sim \kappa q^{2}$. \emph{Therefore,
classical mechanics strongly underestimates the probability of ionization
for small transferred momentum }$q<m_{e}v_{nl}$\emph{.}

The total probability of ionization in classical mechanics is
\begin{equation}
P_{c}(q)=\int_{0}^{\infty }d\kappa \frac{dP_{q}(q,\kappa )}{d\kappa }=\int
\Theta \left( qv_{ex}+\frac{q^{2}}{2m_{e}}-I_{nl}\right) f(\mathbf{v}_{e}%
\mathbf{)dv}_{e}\mathbf{.}  \label{P classical}
\end{equation}%
Equation (\ref{P classical}) simplifies to become%
\begin{equation}
P_{c}(q)=\int \Theta \left( qv_{ex}+\frac{q^{2}}{2m_{e}}-I_{nl}\right)
f_{x}(v_{ex})dv_{ex}.  \label{P classical x}
\end{equation}%
The differential cross section for momentum transfer $q$ is given by
\begin{equation}
d\sigma _{c}(q)=2\pi \rho (q)d\rho (q),  \label{dsigma}
\end{equation}%
where $\rho (q)$ is given by Eq.(\ref{q transfer}). Substituting $\rho (q)$
from Eq.(\ref{q transfer}) into Eq.(\ref{dsigma}) gives%
\begin{equation}
d\sigma _{c}(q)=\frac{8\pi e^{4}Z_{p}^{2}}{v^{2}q^{3}}dq\mathbf{,}
\end{equation}%
which is the Rutherford differential cross section for scattering at small
angles. Finally, the total ionization cross section is
\begin{equation}
\sigma _{c}=8\pi a_{0}^{2}Z_{p}^{2}\frac{v_{0}^{2}}{v^{2}}%
\int_{I_{nl}/v}^{\infty }\frac{P_{c}(q)}{q^{3}}dq\mathbf{.\;}
\label{total ionization classical}
\end{equation}%
In Eq. (\ref{total ionization classical}), we accounted for the fact that
the minimum $q$ is $q=I_{nl}/v$. Note that in the region $q=[1-3]I_{nl}/v$
ionization occurs due the collisions with very fast electrons $v_{e}\sim
v>>v_{nl}$, and $q_{x}\sim q_{z}$. The previous analysis which assumed $%
v_{e}<<v$ and $q_{x}>>q_{z}$ is not valid in this region of extremely small $%
q$. However, because $P_{c}(q)/q^{3}\rightarrow 0$ as $q\rightarrow 0$, this
region of $q=[1-3]I_{nl}/v$ does not contribute to the integral in Eq. (\ref%
{total ionization classical}) and can be neglected. Moreover such small
momentum transfers correspond to very large impact parameter $\rho /v\sim
a_{nl}/v_{nl}$, where the collision becomes adiabatic. Therefore, accurate
calculations yield even smaller $P_{c}(q)$ than in Eq.(\ref{P classical x}).

Equation (\ref{total ionization classical}) is identical to Eq.(\ref{BA
cross section}), where the quantum mechanical ionization probability $%
P_{q}(q,v)$ is replaced by the classical mechanical ionization probability $%
P_{c}(q)$ in Eq.(\ref{P classical x}). The functions $P_{q}(q,v)$ [Eq.(\ref%
{P(q,k)})] and $P_{c}(q)$ [Eq.(\ref{P classical x})] are shown in Fig.8.
Figure 8 shows that the functions $P_{I_{nl}}(q,v)$ and $P_{c}(q)$ are
nearly identical for $q>0.6$. The classical probability of ionization $%
P_{c}(q)$ rapidly tends to zero for $q<0.6$, while the quantum probability
of ionization, $P_{q}(q)\approx 0.283q^{2},$ is much larger than $P_{c}(q)$
at small $q$. The cross section is determined by $P_{q}(q)/q^{3}$. Therefore
the region of small $q$ contributes considerably to the quantum mechanical
cross section. Note that $P_{q}(q)/q^{3}\rightarrow 0$ as $q\rightarrow
I_{nl}/\widetilde{v}$ . It follows that the region of small $q$ contributes
most to the cross section [compare Fig.8(a) for $\widetilde{v}=5$, and
Fig.8(b) for $\widetilde{v}=15$]. For $\widetilde{v}=5$, the classical
mechanical ionization cross section in atomic units is $\sigma _{c}=0.23$,
and the quantum mechanical ionization cross section is $\sigma _{q}=0.30$,
which is 30\% larger than the classical mechanical cross section. For $%
\widetilde{v}=15$, $\sigma _{c}=0.025$ and $\sigma _{q}=0.043$, which is
70\% larger.

\section{Formulary for ionization cross section}

In the high energy limit of fast projectile motion $v>>v_{nl},$ the
classical mechanical calculation can be readily carried out (see Appendix A).

\textbf{The Bohr formula \cite{Thompson}} neglects the electron velocity in
the atom completely, which gives%
\begin{equation}
\sigma ^{Bohr}(v,I_{nl},Z_{p})=2\pi Z_{p}^{2}a_{0}^{2}\frac{v_{0}^{2}E_{0}}{%
v^{2}I_{nl}}.
\end{equation}%
Accounting for the electron velocity gives an additional factor of
$5/3$ compared with the Bohr formula. This gives the classical
mechanical ionization cross section in the limit of high
projectile velocity
\begin{equation*}
\sigma _{classical}^{high\quad energy}(v,I_{nl},Z_{p})=\frac{5}{3}2\pi
Z_{p}^{2}a_{0}^{2}\frac{v_{0}^{2}E_{0}}{v^{2}I_{nl}}.
\end{equation*}%
In the general case with $v\sim v_{nl},$ the classical mechanical
calculation accounting for the finite electron velocity in the atom, but
neglecting the influence of the target nucleus on the electron has been
performed by \textbf{Gerjuoy} \cite{Gerjuoy} [see Appendix A]. This gives%
\begin{equation}
\sigma ^{GGV}(v,I_{nl},Z_{p})=\pi a_{0}^{2}E_{0}^{2}\frac{Z_{p}^{2}}{%
I_{nl}^{2}}G^{GGV}\left( \frac{v}{\sqrt{2I_{nl}/m_{e}}}\right) .
\label{General form formulary}
\end{equation}%
The tabulation of the function $G^{GGV}(x)$\ is presented in Ref.\cite%
{Vriens} for $x>1$, and in Ref.\cite{Armel thesis} for $x<1$, which gives\
\begin{equation}
G^{GGV}(x)=\left\{
\begin{array}{c}
\frac{g(x)}{4x^{2}}\quad for\quad x>1, \\
\frac{0.696}{\exp \left( \frac{0.585-x}{0.096}\right) +1}\quad for\quad x<1%
\end{array}%
\right\} ,  \label{G(x) classical}
\end{equation}%
where%
\begin{equation}
g(x)=\left\{
\begin{array}{c}
\frac{35}{6}+\frac{35}{3\pi }\arctan c+\frac{128\left(
x^{3}b^{3}-b^{3/2}\right) }{9\pi }+\frac{bc}{3\pi }\left( 35-\frac{58b}{3}-%
\frac{8b^{2}}{3}\right) + \\
\frac{2abx}{3\pi }\left[ \left( 5-4x^{2}\right) \left(
3a^{2}+1.5ab+b^{2}\right) -cx\left( 7.5+9a+5b\right) \right] - \\
\frac{16}{\pi }xa^{4}\ln (4x^{2}+1)-ax^{2}\left( 1+\frac{2\arctan c}{\pi }%
\right) \left( 2.5+3a+4a^{2}+8a^{3}\right)%
\end{array}%
\right\} ,  \label{fit G classical}
\end{equation}%
and%
\begin{equation*}
a=1/(1+x^{2})\quad c=3x/4\quad b=1/(1+c^{2}).
\end{equation*}

\textbf{Gryzinski's approximation for the ionization cross section} \cite%
{Gryz} expressed in the form of Eq.(\ref{General form formulary}) is given by%
\begin{equation}
\sigma ^{Gryz}(v,I_{nl},Z_{p})=\pi a_{0}^{2}E_{0}^{2}\frac{Z_{p}^{2}}{%
I_{nl}^{2}}G^{Gryz}\left( \frac{v}{\sqrt{2I_{nl}/m_{e}}}\right) ,
\end{equation}%
where%
\begin{equation}
G^{Gryz}(x)=\left[
\begin{array}{c}
\frac{\alpha ^{3/2}}{x^{2}}\left[ \alpha +\frac{2}{3}(1+\beta )\ln (2.7+x)%
\right] (1-\beta )\left( 1+\beta ^{1+x^{2}}\right) \quad for\quad x>0.206 \\
\frac{4}{15}x^{4}\quad for\quad x<0.206.%
\end{array}%
\right] ,  \label{Gryzinski G(x)}
\end{equation}%
and $\alpha =x^{2}/(1+x^{2})$ $\beta =1/[4x(1+x)]$.

\textbf{Bethe's asymptotic quantum mechanical calculation} \textbf{in the
Born approximation} \cite{Bethe} is valid for $v/v_{0}>2Z_{p}$ and $%
v>>v_{nl} $ \cite{Landau book}, and can be expressed as
\begin{equation}
\sigma ^{Bethe}=4\pi a_{0}^{2}\quad \frac{v_{0}^{4}Z_{p}^{2}}{v^{2}v_{nl}^{2}%
}.\left[ 0.57\ln \left( \frac{v}{v_{nl}}\right) +1.26\right] .
\label{Bethe equation Appendix}
\end{equation}%
The region of validity of the Born approximation and, hence, the Bethe
formula is \cite{Landau book, Bohr}
\begin{subequations}
\begin{equation}
v>\max (2Z_{p}v_{0},v_{nl}).  \label{BA validity appendix}
\end{equation}%
The first condition in Eq.(\ref{BA validity appendix}) assures that the
projectile potential is taken into account in the Born approximation; the
second condition allows use of the unperturbed atomic wave function.

To describe the behavior of the cross section near the maximum, the
second-order correction in the parameter $v_{nl}/v$ has been calculated in
Ref.\cite{Kim}, yielding the cross section in the form
\end{subequations}
\begin{equation}
\sigma _{mod}^{Bethe}(\widetilde{v})=4\pi a_{nl}^{2}\,\frac{v_{0}^{2}}{%
v_{nl}^{2}}\,\frac{Z_{p}^{2}}{\widetilde{v}^{2}}\left[ 0.566\ln \left(
\widetilde{v}\right) +1.26-0.66\frac{1}{\widetilde{v}^{2}}\right] ,
\label{Bethe modified Appendix}
\end{equation}%
where
\begin{equation*}
\widetilde{v}=\frac{v}{v_{nl}}=\frac{v}{\sqrt{2I_{nl}/m_{e}}}%
,\;a_{nl}^{2}=a_{0}^{2}\frac{E_{0}}{2I_{nl}}.
\end{equation*}%
In the general case with $v\sim v_{nl}$, the ionization cross section in the
Born approximation was first calculated in Ref.\cite{Bates}. We have
developed the following fit for the \textbf{Bates and Griffing }result%
\begin{equation}
\sigma _{fit}^{BA}\left( \widetilde{v}=\frac{v}{v_{nl}}\right) =4\pi
a_{nl}^{2}\,\frac{v_{0}^{2}}{v_{nl}^{2}}\,\frac{Z_{p}^{2}}{\widetilde{v}^{2}}%
\left[ 0.283\ln \left( \widetilde{v}^{2}+1\right) +1.26\right] \exp \left[ -%
\frac{1.95}{\widetilde{v}(1+1.2\widetilde{v}^{2})}\right] .
\label{BA our fit for all Formulary}
\end{equation}

\textbf{The Bethe cross section valid for relativistic particles }\cite%
{Bethe book} is given by
\begin{subequations}
\begin{equation}
\sigma _{rel}^{Bethe}=4\pi a_{nl}^{2}\frac{v_{0}^{2}}{v_{nl}^{2}}\frac{%
v_{nl}^{2}Z_{p}^{2}}{v^{2}}\left\{ M_{ion}^{2}\left[ 2\ln \left( \gamma
_{p}\beta _{p}\right) -\beta ^{2}\right] +C_{ion}\right\} ,
\label{Bethe relativistic}
\end{equation}%
where $\beta _{p}^{2}=v_{p}/c$, $c$ is the speed of light, $\gamma _{p}=1/%
\sqrt{1-\beta _{p}^{2}}$, and $M_{ion}^{2}$ and $C_{ion}$ are characteristic
constants depending on the ionized atom or ion. For the hydrogen atom, $%
M_{ion}^{2}=0.283$ and $C_{ion}=4.04$.

\textbf{Gillespie's fit for the ionization cross sections} \cite{Gillespie}
is given by
\end{subequations}
\begin{equation}
\sigma ^{Gill}=\exp \left[ -\lambda _{nl}\left( v_{0}\sqrt{Z_{p}}/v\right)
^{2}\right] \sigma _{mod}^{Bethe},
\end{equation}%
where $\lambda _{nl}$ is a characteristic constant of the ionized atom or
ion (for example, for the ground state of atomic hydrogen, $\lambda
_{nl}=0.76$), and $\sigma _{mod}^{Bethe}$ is the modified Bethe cross
section\ defined in Eq.(\ref{Bethe modified Appendix}).

\textbf{The Olson scaling \cite{Olson} }for the total electron loss cross
section $\sigma ^{el}$, which includes both the charge exchange cross
section $\sigma ^{ce}$ and the ionization cross section, is given by
\begin{equation}
\sigma ^{el}(v,Z_{p})=\pi a_{0}^{2}Z_{p}A_{nl}f^{Olson}\left( \frac{v}{%
v_{0}\gamma _{nl}\sqrt{Z_{p}}}\right) ,
\end{equation}%
where $f(x)$ describes the scaled cross sections%
\begin{equation*}
f^{Olson}(x)=\frac{1}{x^{2}}\left[ 1-\exp \left( -x^{2}\right) \right] ,
\end{equation*}%
and $\gamma _{nl}$ and $A_{nl}$ are constants. For example, $\gamma _{H}=%
\sqrt{5/4}=1.12$ and $A_{H}=16/3$ for atomic hydrogen, whereas $\gamma
_{He}=1.44$ and $A_{he}=3.57$ for helium.

\textbf{Rost and Pattard} \cite{Rand P} proposed a fit for the ionization
cross section, which utilizes two fitting parameters, namely the maximum
value of the cross section and projectile energy corresponding to the
maximum value of the cross section. They showed that if both the cross
section and the projectile velocity are normalized to the values of the
cross section and the projectile velocity at the cross section maximum, then
the scaled cross section $\sigma /\sigma _{\max }$ is well described by the
fitting function \cite{Rand P}
\begin{equation}
\sigma (v)=\sigma _{\max }\frac{\exp (-v_{\max }^{2}/v^{2}+1)}{v^{2}/v_{\max
}^{2}},  \label{R-P fit}
\end{equation}%
where $\sigma _{\max }$ is the maximum cross section, which occurs at the
velocity $v_{\max }$.

We have shown that for ionization by a bare projectile, the values $\sigma
_{\max }$ and $v_{\max }$ are well defined by the projectile charge $Z_{p}$,
with
\begin{eqnarray}
\sigma _{\max } &=&\pi a_{0}^{2}B_{nl}\frac{Z_{p}^{2}}{(Z_{p}+1)}\frac{%
E_{0}^{2}}{I_{nl}^{2}}, \\
v_{\max } &=&v_{nl}\sqrt{Z_{p}+1},
\end{eqnarray}%
where the coefficient $B_{nl}$ depends weakly on the projectile charge. For
example, for ionization of hydrogen by protons, $B_{nl}=0.8$, and for
ionization of hydrogen by bare nuclei of helium or lithium, $B_{nl}=0.93$.

Equation (\ref{R-P fit}) describes well the cross sections at small and
intermediate energies, but underestimates the cross section at high
energies, because it does not reproduce the logarithmic term of the Bethe
formula in Eq.(\ref{Bethe equation Appendix}). To improve the agreement with
the experimental data and the Bethe formula, we propose the new scaling
\begin{equation}
\sigma ^{ion}(v,I_{nl},Z_{p})=\pi a_{0}^{2}\frac{Z_{p}^{2}}{(Z_{p}+1)}\frac{%
E_{0}^{2}}{I_{nl}^{2}}G^{new}\left( \frac{v}{v_{nl}\sqrt{Z_{p}+1}}\right) ,\
\end{equation}%
where%
\begin{equation}
G^{new}(x)=\frac{\exp (-1/x^{2})}{x^{2}}\left[ 1.26+0.283\ln \left(
2x^{2}+25\right) \right] .\;
\end{equation}%
In all previous equations cross section are given per electron in the
orbital. If $N_{nl}$ is the number of electrons in the orbital, the
ionization cross section of any electron in the orbital should be increased
by the factor $N_{nl}$.

Finally, it should be noted that a number of other semi-empirical
models have been developed, which use up to ten fitting parameters
to describe the ionization cross sections over the entire
projectile energy range \cite{Daniel}.


\begin{thebibliography}{99}
\bibitem{HIF reference} B.G. Logan, C.M. Celata, J.W. Kwan, E.P. Lee, M.
Leitner, P.A. Seidl, S.S. Yu, J.J. Barnard, A. Friedman, W.R. Meier, and
R.C. Davidson, Laser and Particle Beams \textbf{20}, 369 (2002).

\bibitem{atmosphere} G.M. Keating and S.W. Bougher, J. Geophys. Res.- Space
Phys. \textbf{97 }(A4), 4189 (1992).

\bibitem{accelerators life time} H. Beyer, V.P. Shevelko (eds), \textit{%
Atomic physics with Heavy Ions} (Springer, Berlin 1999).

\bibitem{spectroscopy} A. Bogaerts, R. Gijbels, and R.J. Carman,
Spectrochimica Acta Part B - Atomic Spectroscopy \textbf{53}, 1679 (1998).

\bibitem{beam stopping} C. Stockl, O. Boine-Frankenheim, M. Geissel, M.
Roth, H. Wetzler, W. Seelig, O. Iwase, P. Spiller, R. Bock, W. Suss, and
D.H.H. Hoffmann, Nucl. Instrum. Meth.A \ \textbf{415}, 558 (1998).

\bibitem{Review atomic physics} S. Datz, G.W. F. Drake, T.F. Galagher, H.
Kleinpoppen, and G. Zu Putlitz, Rev. Mod. Phys. \textbf{71}, S223, (1999).

\bibitem{lenses} P. Chen, Part. Accel. \textbf{20}, 171 (1987); P. Chen,
J.J. Su, T. Katsouleas, S. Qilks, and J. M. Dawson, IEEE Trans. on Plasma
Science PS.-\textbf{15}, 218 (1987).

\bibitem{lenses2} R. Govil, W.P. Leemans, E. Yu. Backhaus and J.S. Wurtele,
Phys. Rev. Lett. \textbf{83}, 3202 (1999).

\bibitem{hep lens} S. Rajagopalan, D.B. Cline, and P. Chen, Nucl. Instrum.
Meth.A \textbf{355}, 169 (1995).

\bibitem{hif plasma focusing} T. Tauschwitz, S.S. Yu, S. Eylon, L. Reginato,
W. Leemans, J.O. Rasmussen, and R.O. Bangerter J. Fusion Engineering and
Design \textbf{32-33}, 493 (1996).

\bibitem{ICF Fast Ignitor} M. Roth, T.E. Cowan, M.H. Key, S.P. Hatchett,
\textit{et al.}, Phys. Rev. Lett. \textbf{86}, 436 (2001); M. Tabak, J.
Hammer, M. E. Glinsky, W. L. Kruer, S. C. Wilks, J. Woodworth, E. M.
Campbell, M. D. Perry, and R. J. Mason, Phys. Plasmas \textbf{1}, 1626
(1994).

\bibitem{Voronov} G. S. Voronov, Atomic Data and Nuclear Data Tables,
\textbf{65}, 1, (1997).

\bibitem{Rudd} M.E. Rudd, Y.-K. Kim, D.H. Madison, and J.W. Galallagher,
Rev. Mod. Phys. \textbf{64}, 441 (1992).

\bibitem{Rudd 2} M.E. Rudd, Y.-K. Kim, D.H. Madison, and T.J. Gay, Rev. Mod.
Phys. \textbf{57}, 965 (1985).

\bibitem{Ogurtsov} G. N. Ogurtsov, Rev. Mod. Phys. \textbf{44}, 1 (1972).

\bibitem{Shvelko book} R.K. Janev, L.P. Presnyakov, V.P. Shevelko, \textit{%
Physics of Highly Charged Ions} (Springer, Berlin 1999).

\bibitem{Dowell1} M.R.C. McDowell and J.P. Coleman, \textit{Introduction to the Theory of Ion-Atom
Collisions} (North-Holland Publishing Company, Amsterdam-London,
1970).

\bibitem{Dowell2}B.H. Bransden and M.R.C. McDowell, \textit{Charge Exchange and the Theory of Ion-
Atom Collisions} (Clarendon Press, Oxford, 1992).

\bibitem{Daniel} E.W. McDaniel, J.B. A. Mitchell and M. E. Rudd, \textit{Atomic Collisions, Heavy
particle Projectiles} ( John Wiley and Sons, Inc., NY, 1993).

\bibitem{Gryz} M. Gryzinski, Phys. Rev.A \textbf{138}, 322 (1965).

\bibitem{webofscience} http://webofscience.com.

\bibitem{Gellipsie} G. Gillespie, J.Phys. B: Mol.Phys. \textbf{15}, L729
(1982); G. Gillespie, Phys. Lett. \textbf{93A}, 327 (1983).

\bibitem{our PoP hif} D. Mueller, L. Grisham, I. Kaganovich, R.L. Watson, V.
Horvat and K.E. Zaharakis, Physics of Plasmas, \textbf{8}, 1753 (2001).

\bibitem{Olson exp} R.E. Olson, R.L. Watson, V. Horvat, and K.E. Zaharakis,
Phys. Rev.A \textbf{67}, 022706 (2003).

\bibitem{Watson exp} R.L. Watson, Y. Peng, V. Horvat, G.J. Kim, and R.E.
Olson, Phys. Rev.A \textbf{67}, 022706 (2003).

\bibitem{Mueller new} D. Mueller, L. Grisham, I. Kaganovich, R. L. Watson,
V. Horvat, K. E. Zaharakis and Y. Peng, Laser and Particle Beams \textbf{20}%
, 551 (2002).

\bibitem{Landau book} L.D. Landau and E.M. Lifshitz, \textit{Quantum
Mechanics} (Addison-Wesley Publishing Co., 1958).

\bibitem{Bohr} N. Bohr, K. Dan. Vidensk. Selsk. Mat.- Fys. Medd. \textbf{18}%
, N8 (1948).

\bibitem{my PAC Xsection} I. D. Kaganovich, E. Startsev and R. C. Davidson,
\textquotedblleft Evaluation of Ionization cross sections in Energetic
Ion-Atom Collisions,\textquotedblright\ Proceedings of the 2001 Particle
Accelerator Conference, (2001).
http://accelconf.web.cern.ch/AccelConf/p01/PAPERS/TPAH314.PDF

\bibitem{my HIF} Igor D. Kaganovich, Edward A. Startsev and Ronald C. Davidson,
Steve R. Kecskemeti, Amitai Bin-Nun, Dennis Mueller and Larry Grisham, Rand L. Watson,
Vladimir Horvat, Konstantinos E. Zaharakis, and Yong Peng,
"Ionization Cross Sections for Ion-Atom Collisions in High Energy Ion Beams",
invited talk at HIF symposium 2004 Princeton, NJ,
to be published in Nuclear Instruments and Methods in Physics Research (2004).

\bibitem{myPRA} I. D. Kaganovich, E. A. Startsev and R. C. Davidson, Phys.
Rev. A \textbf{68}, 022707 (2003).

\bibitem{Japan exp} T. Matsuo, T. Kohno, S. Makino, et al., Phys. Rev. A
\textbf{60}, 3000 (2003).

\bibitem{Thompson} J.J. Thompson, Phil. Mag., \textbf{23}, 449 (1912).

\bibitem{Gerjuoy} E. Gerjuoy, Phys. Rev. A \textbf{148}, 54 (1966).

\bibitem{Ponce} V.H. Ponce, Atomic Data and Nuclear Data Tables, \textbf{19}%
, 63, (1977).

\bibitem{Vriens} L. Vriens, Proc.R. Soc. London, \textbf{90}, 935 (1966).

\bibitem{Armel thesis} M. Scott Armel, Ph.D. Thesis, University of
California at Berkeley (2000);
http://faculty.oxy.edu/scottfunk/Sci/index.html.

\bibitem{Bethe} H. Bethe, Ann. Phys. (Leipz.) 5, 325 (1930).

\bibitem{Bethe book} H.A. Bethe and R. Jackiw, \textit{Intermidiate Quantum
Mechanics} (The Benjamin/Cummings Publishing Company, sec.ed., 1968).

\bibitem{Shah} M. B. Shah, D..S. Elliott and H. B. Gilbody, J.Phys. B:
Mol.Phys. \textbf{20}, 2481 (1987).

\bibitem{red books} Atomic Data for Fusion. Volume 1: Collisions of H, H2,
He and Li Atoms and Ions with Atoms and Molecules (C. F. Barnett
ed.) ORNL-6086 (1990). Atomic Data for Fusion. Volume 5:
Collisions of Carbon and Oxygen Ions with Electrons, H, H2 and He
(R. A. Phaneuf, R. K. Janev, M. S. Pindzola) ORNL-6090 (1987);
http://www-cfadc.phy.ornl.gov/redbooks/. Note that data for carbon
ions are semiemphirical. Only one data point at 400keV/amu exists
and the rest of the data were reproduced using Gillespie's fit.

\bibitem{Shah Li atoms} M. B. Shah and H. B. Gilbody, J. Phys. B: Mol. Phys.
\textbf{15}, 413 (1982).

\bibitem{Alice paper} A. Kolakowska, M. S. Pindzola, and D. R. Schultz,
Phys. Rev. A \textbf{59}, 3588 (1999).

\bibitem{Kim} Y.K. Kim and M. Inokuti, Phys. Rev. A \textbf{3}, 665 (1971);
M. Inokuti, Rev. Mod. Phys. \textbf{43}, 297 (1971).

\bibitem{Shevelko paper} V.P. Shevelko, I. Yu. Tolstikhina and Th.
Stoehlker, Nucl. Instr. Meth. B \textbf{184}, 295 (2001).

\bibitem{Gillespie} G.H. Gillespie, Phys. Rev. A \textbf{18}, 1967 (1978).

\bibitem{Bates} D. R. Bates and G. Griffing, Proc.Phys. Soc. London, \textbf{%
66}, 961 (1953).

\bibitem{Bohr Linhard} N. Bohr and J. Linhard, K. Dan. Vidensk. Selsk. Mat.-
Fys. Medd. \textbf{28}, 1 (1954).

\bibitem{Olson} R.E. Olson, Phys. Rev. A \textbf{18}, 2464 (1978).

\bibitem{Janev} R.K. Janev, Phys. Rev. A \textbf{18}, 1810 (1983).

\bibitem{Stolterfoht} N. Stolterfoht, R.D. DuBois and R.D. Rivarola, \textit{%
Electron Emission in Heavy Ion-Atom Collisions} (Springer, 1997).

\bibitem{Rand P} J.M Rost and T. Pattard, Phys. Rev. A \textbf{55}, R5
(1996).

\bibitem{Shah He 85} M. B. Shah and H. B. Gilbody, J. Phys. B: Mol. Phys.
\textbf{18}, 899 (1985).

\bibitem{Shah He 89} M. B. Shah, P. Mc.Callion and H. B. Gilbody, J. Phys.
B: Mol. Phys. \textbf{22}, 3037 (1989).

\bibitem{expC6}A.S. Schlachter, K.H. Berkner, W.G. Graham, et al., Phys. Rev. A \textbf{23}, 2331
(1981).

\bibitem{expIandU} S. Datz, R. Hippler, L.H. Andersen, et al., Phys. Rev. A \textbf{41}, 3559 (1990).

\bibitem{expAu} H. Berg, J. Ullrich, E. Bernstein, J. Phys. B: Mol. Phys. \textbf{25}, 3655 (1992).

\bibitem{He theory} P.D. Fainstein, V.H. Ponce, and R. D. Rivarola, J.Phys.
B: Mol. Phys. \textbf{24}, 3091 (1991).

\bibitem{q2scaling} , H.K. Haugen, L.H. Andersen, P. Hvelplund, and H.
Knudsen, Phys. Rev. A \textbf{26}, 1950 (1982).

\bibitem{Theod1} C.E. Theodosiou, Phys. Rev. A \textbf{38}, 4923 (1988).

\bibitem{Theod2} C.E. Theodosiou, Phys. Rev. A \textbf{36}, 2067 (1987).

\bibitem{Theod3} W. Sander and C.E. Theodosiou, Phys. Rev. A \textbf{42},
5208 (1990).

\bibitem{McGuire} J.H. McGuire, Phys. Rev. A \textbf{26}, 143 (1982).

\bibitem{Duman} E.L. Duman, L.I. Men'shikov, and B.M. Smirnov,
Sov.-Phys.-JETP \textbf{49}, 260 (1979).

\bibitem{Matveev} V.I. Matveev, V.A. Pazdzerkii, and Kh. Yu. Rakhimov,
Technical Physics \textbf{46}, 512 (2001).

\bibitem{Olson vhalf} R.E. Olson, Phys. Rev. A \textbf{27}, 1871 (1983).

\bibitem{Ovchnnikov} S.Y. Ovchinnikov, Phys. Rev. A \textbf{42}, 3865 (1990).

\bibitem{Abramov} D.I. Abramov, S.Y. Ovchinnikov, E.A. Solov'ev, Phys. Rev.
A \textbf{42}, 6366 (1990).

\bibitem{Ovchnnikov PRL} S.Y. Ovchinnikov and J.H. Macek, Phys. Rev. Lett.
\textbf{75}, 2474 (1995).

\bibitem{Shah 2000} B.S. Nesbitt, M.B. Shah, SFCO' Rourke, et al., J.Phys. B: Mol. Phys. \textbf{33}, 637
(2000).


\end{thebibliography}
\end{document}